# Numerical analysis of dilute methanol spray flames in vitiated coflow using extended Flamelet Generated Manifold model

Bharat Bhatia,[1] Ashoke De,[1,2,a)] Dirk Roekaerts[3] and Assaad R. Masri[4]

[1]*Department of Aerospace Engineering, Indian Institute of Technology Kanpur, 208016, Kanpur, India.*

[2]*Department of Sustainable Energy Engineering, Indian Institute of Technology Kanpur, 208016, Kanpur, India.*

[3]*Department of Process and Energy, Delft University of Technology, 2628 CB Delft, The Netherlands.*

[4]*School of Aerospace, Mechanical, and Mechatronic Engineering, The University of Sydney, Sydney, NSW 2006, Australia.*

The present work focuses on the large eddy simulation (LES) and the study of turbulent dilute methanol spray flames in vitiated coflow using the secondary-oxidizer Flamelet Generated Model (FGM). The modified FGM model uses an additional secondary oxidizer parameter in addition to the three other parameters previously used for spray flames – progress variable, mixture fraction, and enthalpy. The results for gas phase and droplet properties are validated against the dilute methanol spray flame database for varying fuel injection amounts. The droplet statistics and the lift-off flame heights are accurately captured for all the cases. A proper orthogonal decomposition (POD) of the scalar fluctuating hydroxyl radical (OH) field and velocity-temperature field capture the flame structures in the downstream region of ignition kernels. The detailed POD analysis reveals that the base frequency of the dominant OH field equals that of the dominant vortical structure of 67.3 Hz. The flame propagation happens around these dominant vortical structures because of the less-strained fluid mixing.

## I. INTRODUCTION

Multiphase combustion systems form the backbone of many industrial processes, especially power generation devices which commonly involve combustion of the liquid fuels and conversion of their chemical energy to mechanical energy. This energy conversion chain from chemical to thermal and thermal to mechanical requires efficient combustion with the least pollution. An injected liquid fuel is first atomized into fine droplets to undergo easy and quicker evaporation, followed by efficient mixing with the oxidizer before combustion. To improve the traditional combustion mode, researchers have considered an alternative mode referred to as "flameless combustion," which has several variants radically different from the traditional combustion of pure air and fuel at room temperature[1-3]. The heating of both fuel and oxidizer before combustion in High-temperature Combustion Technology (HiCOT), preheating of air only in High-Temperature Air Combustion (HiTAC)[4], varying the oxygen content in air through dilution in Moderate or Intense Low Dilution (MILD)[5] or Colorless Distributed Combustion (CDC)[6-13] are few examples of these new combustion systems.

___

a) Electronic mail: ashoke@iitk.ac.in.

A high-temperature diluted oxidizer has been used to study partially premixed flames for vitiated-coflow gaseous flames[14] or auto-igniting dilute spray flames[15]. In the case of spray flames, the evaporation process and the auto-ignition are critical aspects in determining flame type and location. The capturing of ignition and extinction processes is needed for better predictions of flame structure and the involved intermediate species. The auto-ignition is a delicate play between small-scale turbulent structures, chemical reactions, and fuel evaporation. The ignition process has a direct effect on the lift-off height[16]. It was shown in Ref. 17 that the reaction rates and ignition kernels are not as strongly related to the scalar dissipation rate as in the gaseous flames. Nevertheless, auto-igniting flames are linked to lower values of scalar dissipation rate.

The dilute methanol spray burner with a vitiated coflow[15] provides detailed data on the auto-igniting flame. Several modeling approaches have already been applied to these flames. Prasad et al.[18] used the transported-pdf method with Eulerian stochastic fields. It was found that the ignition kernels develop in the region of lean mixtures where the mixture may be termed "the most reactive mixture" and finally lead to a fully developed flame. Large-eddy simulation (LES) has been performed using the transported-pdf approach with stochastic Lagrangian particle models for the combustion process and the spray droplet tracking[19]. Also, MMC-LES, a combination of PDF and CMC, has been used in Ref. 20.

Further modeling options are models using mean species transport equations with a source term closure based on a micro-reactor model, for example, EDC models. They have been used to study the combustion and the ignition characteristics in jet-in-hot-coflow (JHC) burners[21-23]. A disadvantage of EDC is that many species equations have to be solved in the turbulent flame calculation. The flamelet methods or generally 'tabulated chemistry models' are computationally more efficient and potentially more accurate where the chemical evolution is projected in a lower-dimensional manifold described by a few independent scalar variables. In this article, we elaborate on a model of this category, namely a flamelet generated manifold (FGM) model. It describes the chemical evolution in turbulent flames via a small number of independent scalar variables. The source terms for these variables are pre-computed from a relevant set of laminar flames. This method can also cover regimes with flame extinction and re-ignition and has been validated systematically for many cases[24,25]. It also has been extended to describe flameless combustion. But a distinction should be made between jet in vitiated coflow systems and furnaces. In the former case, the dilution is present in the inflow; in the latter, the diluent is produced in the primary flame and recirculated aerodynamically. Both instances require different generalizations of FGM. In the case of vitiated coflow with uniform composition, a description using a single mixture fraction and a progress variable (and possibly enthalpy) is sufficient. When the dilution of the coflow is nonuniform, a second mixture fraction can be added, as was done by Sarras et al.[26]. In the case of dilution by internal recirculation, the dilution level is generated dynamically, and an additional parameter can be introduced to describe it, as was done in the DAFGM presented by Huang et al.[27]. This model can also be applied to jet in vitiated coflow burners where the main flame and the secondary burner employ the same fuel. In the dilute methanol auto-igniting flame used in the present study, the hot coflow is generated from a $H_2$/air flame and does not contain the product of the main flame, thus rendering the DAFGM[27] inapplicable. Instead, in the current study, the Flamelet Generated model (FGM) is extended with an additional parameter to distinguish between the two oxidizers with different compositions and temperatures. The FGM structure is similar to that developed by Sarras et al.[26]. The model is validated against the auto-igniting methanol flame as a test case. The availability of extensive data related to droplet statistics and flame structure allows for a complete test. However, the relation between flow properties and chemical reaction is not very clear from single-point statistics. To provide deeper insight into the reacting flow, we perform a proper orthogonal decomposition (POD) on the velocity-temperature field and the OH field of the flame to identify the dominant fluctuating structures. These structures help us find the ignition kernels and the flame growth locations that, otherwise, are difficult to locate in an instantaneous field. The procedure allows seeing relations between mixture fraction fields, vortical structures (fluid strain), and the creation of a flame, using the OH species as a marker for flame structure[28-30].

## II. NUMERICAL METHODOLOGY

The multiphase simulations are carried out using an Eulerian-Lagrangian approach. The dispersed (liquid) phase is solved with the Lagrangian approach, while the continuous phase is solved with the Eulerian framework. The gas-phase reaction of evaporated fuel and gaseous oxidizer is modeled with the FGM (Flamelet Generated Manifold) model. Eulerian equations, Lagrangian equations, and FGM are explained in the following.

### A. Basic Governing Equations

Filtered transport equations are solved for the Eulerian (gas) phase, continuity equation and momentum equation, energy equation, and scalar equations related to the FGM model. Continuity and momentum equations, including sources describing transfer between the two phases, are given by:

$$\frac{\partial \overline{\rho}}{\partial t} + \nabla(\overline{\rho}\widetilde{u}) = \overline{S_{mass}} \qquad (1)$$

$$\frac{\partial \overline{\rho}\widetilde{u}_i}{\partial t} + \frac{\partial}{\partial x_j}(\overline{\rho}\widetilde{u}_i\widetilde{u}_j) = -\frac{\partial \overline{p}}{\partial x_i} + \overline{\rho}g_i + \frac{\partial}{\partial x_j}\overline{\tau}_{ij} + \overline{S_{mom,i}} \qquad (2)$$

The term $\overline{\tau}_{ij}$ in Eq. 2 denotes the stress tensor, including laminar and subgrid-scale effects. The unresolved stress is closed by the dynamic k-equation model, which is a one-equation eddy viscosity SGS model. This model was primarily developed by Kim & Menon[31] based on the previous works of Germano et al.[32]. In recent years, many improvements to the models have been proposed[33-34].

$$\overline{\tau_{ij}} = 2\overline{\rho}\nu_{eff}\left(S_{ij} - \frac{1}{3}\delta_{ij}S_{ii}\right), \tag{3}$$

$$S_{ij} = \frac{1}{2}\left(\frac{\partial u_i}{\partial x_j} + \frac{\partial u_j}{\partial x_i}\right) \tag{4}$$

and $\nu_{eff}$ is the effective kinematic viscosity (both molecular and sub-grid viscosity). The turbulent kinetic energy is obtained from

$$\frac{\partial k_{sgs}}{\partial t} + \frac{\partial \widetilde{U}_j k_{sgs}}{\partial x_j} = \frac{\partial}{\partial x_j}\left(\nu_t \frac{\partial k_{sgs}}{\partial t}\right) - C_e \frac{k_{sgs}^{3/2}}{\Delta} - 2\nu_{sgs}\widetilde{S}_{ij}\widetilde{S}_{ij} \tag{5}$$

$$\nu_{sgs} = C_k \sqrt{k_{sgs}}\Delta \tag{6}$$

where $k_{sgs}$ is the sub-grid scale kinetic energy. The $C_k$, $C_e$ constants are computed based on the dynamic formulation from Germano et al.[32] Mean scalar equations (enthalpy and the scalars of the FGM model) are closed using the gradient diffusion assumption for the unresolved flux. The partial differential equations (PDEs) are discretized using a Finite Volume Method (FVM) code implemented in OpenFOAM-v19.0[35]. The convective flux discretization uses a second-order TVD (Total Variation Diminishing) scheme. In contrast, viscous flux discretization involves a second-order central scheme. The temporal term deploys the second-order Crank-Nicolson scheme with sufficiently small time steps to maintain stability and reduce numerical diffusion.

## B. Disperse Phase Modelling

### 1. Lagrangian Particle Tracking (LPT)

The Lagrangian particles represent the liquid droplets, which act as a point source of mass, momentum, and energy for the above equations. In the numerical approach, the full spray is represented by parcels, and computational particles represent a set of physical droplets with identical properties. In the following discussion, they will be denoted as particles or droplets. The equations for particles contain sub-models for dispersion, collision, atomization, and heat transfer. The Basset-Boussinesq-Ossen (BBO) equation is solved for momentum conservation of the Lagrangian particles. The equation is given by,

$$\frac{d\mathbf{x}_p}{dt} = \mathbf{u}_p \qquad (7)$$

$$m_p \frac{d\mathbf{u}_p}{dt} = \mathbf{F}_D + \mathbf{F}_G + \mathbf{F}_T \qquad (8)$$

where $\mathbf{x}_P$, $m_p$, $\mathbf{u}_P$ is each particle's position, mass, and velocity, respectively. $\mathbf{F}_D$ and $\mathbf{F}_G$ are the drag and gravitational forces (body forces) acting on the particles, whereas $\mathbf{F}_T$ is the turbulence effect modeled using gradient dispersion model. They are calculated as follows:

$$\mathbf{F}_D = C_D \frac{\pi D_p^2}{8} \rho_g (\tilde{\mathbf{u}} - \mathbf{u}_p) |\tilde{\mathbf{u}} - \mathbf{u}_p| \qquad (9)$$

$$\mathbf{F}_G = m_p g \left(1 - \frac{\overline{\rho}}{\rho_p}\right) \qquad (10)$$

$\mathbf{F}_G$ accounts for both gravity and buoyancy effects, $D_p$ is the diameter of the droplet, $\tilde{\mathbf{u}}$ and $\mathbf{u}_p$ are the filtered velocity of gas-phase at the droplet location and the velocity of the droplet, respectively. $C_D$ is the drag coefficient of droplets, which is calculated from the Schiller-Naumann equation[36]:

$$C_D = \begin{cases} 24(1 + 0.15 Re_p^{0.687})/Re_p & Re_p \leq 1000 \\ 0.44 & Re_p > 1000 \end{cases} \qquad (11)$$

$$Re_p = \frac{\rho_g |\tilde{\mathbf{u}} - \mathbf{u}_p| D_p}{\mu_g} \qquad (12)$$

where $Re_p$ is the particle Reynolds number and $\mu_g$ is the gas phase dynamic viscosity. The KHRT (Kelvin-Helmholtz and Rayleigh-Taylor) model describes the aerodynamic break-up for secondary atomization. Su et al.[37] showed that the KHRT model predicts a better drop size distribution, closer to the experimental results than the KH model alone. Later, Ricart et al.[38] used the KHRT model with a break length concept in which the KH instability dominates the breakup in the area within breakup length. Outside the breakup length in the region of secondary atomization, RT instability starts to compete with the KH instability. In other words, the breakup length concept, along with the KH-RT model, considers both the primary atomization and the secondary atomization. Since the present case simulates the dilute sprays, only the secondary atomization is dominant. Thus, the KH-RT breakup of the droplets occurs without considering the breakup length concept. Hence, the model by Su et al.[37] is used in the present study. The wavelength of the KH waves on the droplet surface and the characteristic breakup time is calculated as,

$$\Lambda_{KH} = \frac{9.02r\left(1+0.45\sqrt{Z}\right)\left(1+0.4T^{0.7}\right)}{\left(1+0.865We_g^{1.67}\right)^{0.6}} \quad (13a)$$

$$\tau_{KH} = \frac{3.726B_1 r}{\Omega_{KH}\Lambda_{KH}} \quad (13b)$$

Here, $r$ is the droplet radius, $We_g$ is the gas Weber number, Z is the Ohnesorge number, $T(=Z\sqrt{We_g})$ is the Taylor number, and $\Omega_{KH}$ is the frequency of the KH wave. The wavenumber of the RT waves on the droplet surface and the characteristic break time is calculated as,

$$K_{RT} = \sqrt{\frac{-g_t\left(\rho_f - \rho_a\right)}{3\sigma}} \quad (14a)$$

$$\tau_{RT} = \frac{C_\tau}{\Omega_{RT}} \quad (14b)$$

The acceleration $g_t$ is in the direction of the droplet trajectory, $C_\tau$ is a constant assumed equal to unity, and $\Omega_{KH}$ is the frequency of the RT wave[39]. Assuming spherical droplets, the standard Ranz-Marshal[40] and Frossling correlations[41] are used for convective heat transfer and mass transfer.

$$Nu = 2 + 0.552 Re_p^{0.5} Pr^{0.33} \quad (15)$$

$$Sh = 2 + 0.552 Re_p^{0.5} Sc^{0.33} \quad (16)$$

To take into account the blowing effect due to the droplet evaporation resulting in thickening of the laminar boundary layer and reduced transfer rate, the Sherwood ($Sh$) and Nusselt numbers ($Nu$) are replaced by modified values (denoted by *) following Abramzon and Sirignano[42]

$$Sh^* = 2 + \frac{Sh-2}{F_M}, \qquad Nu^* = 2 + \frac{Nu-2}{F_T} \qquad (17)$$

The $F_M$ and $F_T$ are the same universal function of the corresponding transfer numbers denoted by,

$$F = (1+B)^{0.7} \frac{\ln(1+B)}{B}. \qquad (18)$$

The blowing effect becomes significant at higher temperatures. In the case of temperatures reaching above boiling point, the flashing of liquid is also evident. Thus, a combination of evaporation and the flashing process is based on the model proposed by Zuo et al.[43].

$$\dot{m}_e = \frac{\pi k d_0}{c_p} \left( \frac{Nu^*}{1+\dot{m}_f/\dot{m}_e} \right) \ln\left( 1 + \left(1 + \frac{\dot{m}_f}{\dot{m}_e}\right) \frac{h_\infty - h_b}{h_{fg}} \right). \qquad (19)$$

Here $\dot{m}_e$ is the evaporation rate, which depends on the heat conductivity $k$, heat capacity $c_p$, and latent heat of vaporization $h_{fg}$. Also, $h_\infty$ and $h_b$ are the enthalpy of gas in the gas phase and at the droplet surface. The NSRDS – AICHE (National Standard Reference Data System - American Institute of Chemical Engineers) database of National Institute Standards and Technology[44] is used to calculate the thermodynamic properties for the two phases at different pressures and temperatures.

**C. Flamelet generation**

In the FGM method, the local states in a turbulent flame are described using a few controlling variables, either representing the mixing (mixture fraction) or the chemical evolution (progress variables). The other composition variables are related to them according to relations found in a selected set of one-dimensional laminar flames known as *flamelet*[45]. In the first step, a group of flamelets is pre-computed, and the structure information is stored in a library. Assuming that the flamelet relations hold in the turbulent flame, only the dynamics of the controlling variables have to be solved. But in RANS and LES, it has to be taken into account that the controlling variable has turbulent fluctuations. In the present study, the code CHEM1D[46] is employed for the flamelets' calculation at different conditions. It solves one-dimensional temperature, species, and flow equations with an equation of state (EOS) using adaptive grid refinement to define the one-dimensional flame structure completely[47]. For the present atmospheric flames, the ideal gas EOS has been used. The states closest to the chemical equilibrium state are reached when the scalar dissipation value (or strain rate) approaches zero. These are the flamelets with the maximum temperature near the adiabatic mean temperature depicted in Fig. 1. On the other hand, the mixing solution is obtained without reaction.

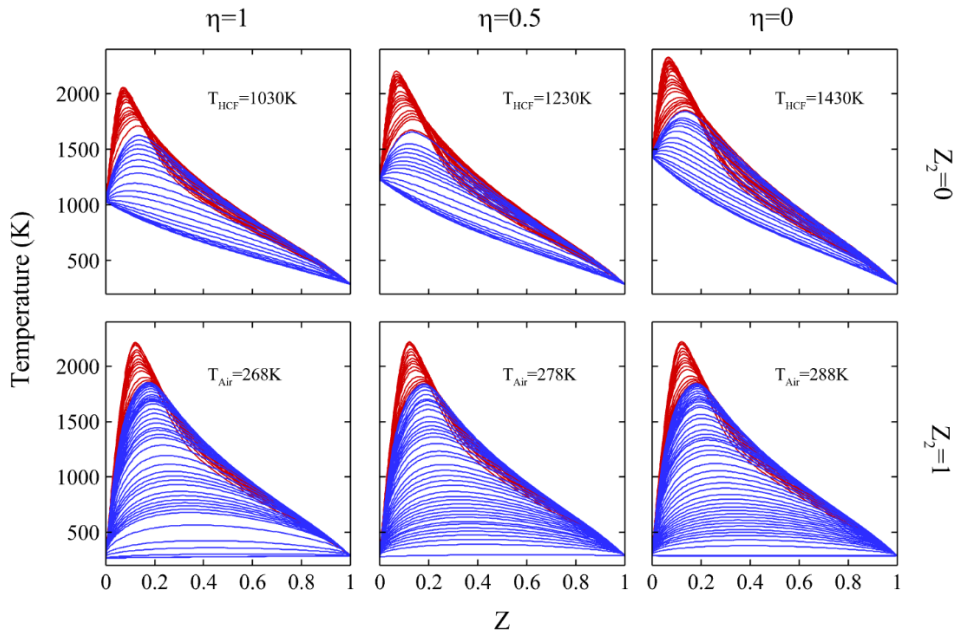

Fig. 1. Flamelet relations for temperature as a function of mixture fraction at different dissipation rates for Mt2B and Mt2C flames. Steady-state flamelets are shown in red, whereas the unsteady extinguishing flamelet states are shown in blue.

Fig. 1 shows the flamelet relations between temperature and mixture fraction for methanol as a fuel and air or hot coflow as an oxidizer in counter-flow configuration for 32 species, 167 reactions chemical mechanism[48]. The flamelets correspond to different oxidizer temperatures to account for the fuel evaporation, the details of which are specified in the following sub-section. At a given oxidizer temperature, the flamelets are created for a complete range of dissipation rates from very small to the extinction value. As visible in Fig. 1, the red-colored lines are steady flamelets available for scalar dissipation rates lower than the critical value above which the flamelets move towards the mixing line (extinguish). The range of steady flamelets varies for each set of conditions; for example – the steady flamelets for $Z_2=1$ and $\eta=1$ range from near zero to around 400s$^{-1}$ of strain rate, whereas steady flamelets for $Z_2=0$ and $\eta=1$ range from near zero to around 3000s$^{-1}$ of strain rate. The FGM generated from flamelets needs a complete range of progress variable values, including the states between the mixing line and the steady flamelet with the highest strain rate. The states in that region can be generated in different ways. A first method considers a transient flamelet at a scalar dissipation rate above the critical value. The FGM library created from flamelets generated by this methodology is called *Extinguishing FGM* (EFGM). The blue-colored lines in Fig. 1 are states of the extinguishing flamelet. Alternatively, when at a relevant scalar dissipation rate, the mixing solution auto-ignites the transient states reached during ignition and can be used to construct the manifold. The corresponding FGM is called *Igniting FGM* (IFGM). IFGM can be used in the case of a hot vitiated coflow (hot oxidizer). In the system studied here, cold air acts as an oxidizer, and a treatment using only IFGM is impossible, and EFGM is used.

**D. Flamelet library generation**

The flamelets described in the previous section correspond to an oxidizer boundary condition of hot vitiated coflow and do not consider heat loss. To obtain a complete library required for the turbulent flame generation, the flamelet simulations have to be repeated, varying the oxidizer composition while also including the heat loss. This leads to a library with four independent degrees of freedom, as explained in the following. The methanol spray flame case has two oxidizer streams: an airflow from the central jet core maintained at room temperature and hot coflow produced by lean H2/Air flames. Both streams have a constant but different temperature. This system can be handled straightforwardly by introducing an extra mixing parameter (mixture fraction) to define the relative fraction of the two oxidizer streams in the three-stream mixture with fuel. It should be noted that the system does not fit the scope of models considering a product stream as a third stream, either diluted-air-based FGM models[27] or flamelet progress variable models[49]. Indeed, the vitiated coflow is different in composition from the product stream of the main flow.

### 1. Mixture Fraction ($Z_1$)

The primary mixture fraction ($Z_1$) represents the mass fraction of the fuel in a mixture of fuel-oxidizer. By definition in a two-stream problem the composition of the oxidizer stream and the fuel stream respectively correspond to mixture fraction values 0 and 1 at progress variable 0. With $\psi$ denoting any relevant thermochemical variable needed to describe the state of the mixture, the boundary condition is

$$\psi(Z_1 = 0, C = 0) = \psi^{Ox} \tag{20}$$

$$\psi(Z_1 = 1, C = 0) = \psi^{F} \tag{21}$$

### 2. Oxidizer mixture fraction ($Z_2$)

The central carrier jet consists of cold air at 283-288K, and some evaporated fuel. The hot outer coflow is the product of hydrogen/air flame burnt in lean conditions at an approximately constant temperature of 1430K. Further, the product of lean combustion of premixed hydrogen/air flame majorly consists of $H_2O$, $N_2$, and $O_2$ only. The difference between composition and temperature demands a separate parameter that tracks both the oxidizers' mixing and helps in better definition of thermochemical quantities. Sarras et al.[26] used a similar parameter to define and distinguish the air from vitiated coflow of varying oxygen content in Delft Jet-in-Hot-Coflow (DJHC). This methodology also allows for flamelet generation for oxidizers at intermediate states of oxygen content. It should be noticed that the current case cannot be treated as MILD combustion with exhaust gas recirculation (EGR). In EGR, fuel combustion products with the oxidizer act as an inert diluent to the oxidizer. There are various studies on EGR-based flames like furnace flames[27,49]. Furthermore, the diluent temperature will be the adiabatic flame temperature from the complete combustion of fuel and oxidizer, provided the enthalpy loss is neglected. If air is considered the oxidizer in the present case, the combustion product will contain a carbon-based product like $CO_2$, which is absent in the hot coflow. In other words, the temperature of 1430K in hot coflow will require a completely different hot-coflow composition if the coflow was to be made by diluting the air with exhaust gas. This justifies the treatment of cold air and hot coflow as two separate oxidizers for the present case. Moreover, the oxidizer mixture fraction helps to accurately define the amount of fuel evaporation in the carrier jet. The oxidizer mixture fraction ($Z_2$) is a normalized scalar representing a mixture of two oxidizers – cold air and hot coflow. By definition, the composition of the hot coflow stream and the air respectively correspond to second mixture fraction values 0 and 1.

$$\psi^{Ox}(Z_2 = 0) = \psi^{HCF} \tag{22}$$

$$\psi^{Ox}(Z_2 = 1) = \psi^{Air} \tag{23}$$

Flamelets can be generated for both the oxidizer streams or any mixture of the two oxidizer streams by varying the second mixture fraction.

### *3. Scaled Progress Variable (C)*

The set of computed flamelets can be tabulated as a function of the two mixture fractions, a progress variable tracking the conversion progress, and a variable characterizing the enthalpy deficit relative to an adiabatic reference state. The progress variable is scaled to a range [0,1] for tabulation convenience. The value of *C* equals zero would define either a pure state of fuel or oxidizer (as shown in Eq. 20,21) or a state of mixing between them[50]. Figure 2 shows the temperature as a function of scaled progress variable and mixture fraction at different values of enthalpy deficit ($\eta$) and $Z_2$. The brighter white region indicating high temperature lies near the top, where the value of *C* is close to 1. At the same time, C equals zero means the temperature of a nonreactive mixing of oxidizer and fuel. The high flame temperature zone decreases as the enthalpy deficit or second mixture fraction increases. Reducing the second mixture fraction represents a mixture containing more hot-coflow and a higher oxidizer temperature. This is why a broader high-temperature zone arises for $Z_2$ equals 0, and it results in quicker auto-ignition of the fuel-oxidizer mixture even in lean conditions at lower mixture fraction values.

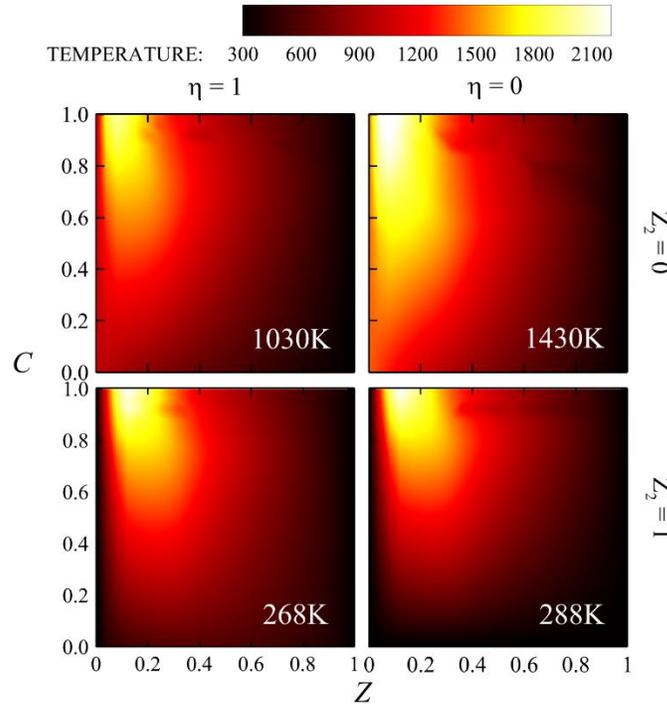

FIG. 2. Temperature variation as a function of reaction progress variable and mixture fraction in the FGM for Mt2B and Mt2C flames. The flamelets are shown only for extreme values 0 and 1 of enthalpy deficit $\eta$ and second mixture fraction $Z_2$.

### *4. Enthalpy deficit ($\eta$) or non-adiabatic FGM*

The multiphase reactive flow, considered here, involves liquid fuel evaporation before mixing and combustion. The evaporation process reduces the local temperature and enthalpy of the surrounding fluid, affecting the reaction zone and flame structure. Higher enthalpy losses due to evaporation may also lead to local flame extinction. It becomes more critical for the auto-igniting flames where the formation of ignition kernel is decisive for flame lift-off height. The region near to fuel injection nozzle has a high density of spray droplets, and the liquid fuel evaporation delays the auto-ignition, thus, leading to an increase in the lift-off height. Hence, the enthalpy deficit represents the difference between the actual condition and the adiabatic case. Similarly, a cold wall may also lead to enthalpy loss, local quenching, and detachment of a flame. Different methods for including heat loss in a flamelet library have been used in earlier studies. The first would be to calculate non-adiabatic flamelets by including a heat loss term in the energy equation[51]. Since the phenomena of heat loss in a single flamelet and a spray flame environment are quite different, the inclusion of enthalpy loss due to evaporation in the flamelet library will not cover a correct range of conditions[52]. Another method to achieve enthalpy deficits is to generate flamelets for reduced temperature boundary conditions. Marracino & Lentini[53] assumed an equal enthalpy loss for both the fuel and oxidizer in creating flamelets. In contrast, Ma et al.[24-25], and Huang et al.[27] assumed the enthalpy loss for the oxidizer only. The former examines a methane/air flame for radiation effects, thus giving rise to source/sink terms in the enthalpy equation. The latter is concerned with the multiphase reactive system where a part of a liquid fuel droplet is assumed to absorb the latent energy from its surrounding (oxidizer) during the evaporation. It results in a more pronounced effect on the oxidizer only. The latter approximation is adopted for enthalpy loss effects in the present study. Hence, the enthalpy deficit ($\eta$) condition would apply to the oxidizer only, and the flamelet state at the boundary is given by:

$$\psi(Z_1 = 0, C = 0) = \psi^{Ox}(Z_2, \eta). \qquad (24)$$

The value of the maximum enthalpy deficit must be specified depending on the insight of the studied flame. The magnitude of the considered range of enthalpy deficit is temperature and composition-dependent and therefore depends on $Z_2$. But for every case, the range is described by a parameter $\eta$ with range [0,1].

**E. Extended FGM Model**

*1. Construction of a lookup FGM table*

The thermochemical states of flamelets calculated in Section II(C) can be written as functions of the scaled independent variables $\psi(Z_1, C, \eta, Z_2)$. The range of tabulated conditions depends on the set of flamelets considered and should be sufficiently large. The definition of two mixture fractions defines all possible mixing states. The set of flamelets is well complete to cover the entire conversion process from inert to fully burnt state. The heat loss range is also sufficient to cover the actual heat loss due to evaporation. In a turbulent flame, the independent variables are fluctuation. In LES, subgrid-scale fluctuations are not resolved and, instead, they are modeled using a presumed joint PDF of the independent variables. The density-weighted filtered thermochemical quantities are obtained from the flamelet data by integrating over the PDF[50]:

$$\tilde{\psi} = \int_0^1 \int_0^1 \int_0^1 \int_0^1 \psi(Z_1, C, \eta, Z_2) \tilde{P}(Z_1, C, \eta, Z_2) dZ_1 dC d\eta dZ_2 \qquad (25)$$

It is assumed that all the control variables are independent of each other. Furthermore, the mixture fraction $Z_1$ and scaled progress variable C are assumed to vary according to the $\beta$-function PDF[54], entirely determined by the mean and the variance of the considered variable. The fluctuations in enthalpy deficit and second mixture fraction can be neglected (modeled by a Dirac $\delta$-function PDF). Fluctuations in enthalpy deficit are expected to have a negligible impact because the variation of properties with deviation from the mean deficit is monotonic and very smooth. Fluctuations of the second mixture fraction are expected to have much less impact than fluctuations in the first mixture fraction because it only describes a variation in oxidizer composition, whereas the first mixture fraction describes the mixing between fuel and the combination of both oxidizers and controls the auto-ignition. Then, Eq. (25) may be rewritten as:

$$\tilde{\psi} = \int_0^1 \int_0^1 \psi(Z_1, C, \eta, Z_2) \tilde{P}(Z_1) \tilde{P}(C) dZ_1 dC \tag{26}$$

Filtered properties obtained by pre-integration over the control variable PDF are stored in a six-dimensional *lookup* table for the flow simulations:

$$\tilde{\psi} = \psi\left(\widetilde{Z_1}, \widetilde{\zeta_{Z1}}, \widetilde{C}, \widetilde{\zeta_C}, \tilde{\eta}, \widetilde{Z_2}\right) \tag{27}$$

The independent variables of the table are the resolved values of the four control variables of the laminar table and the scaled variances of $Z_1$ and $C$, which are denoted by $\widetilde{\zeta_{Z1}}$ and $\widetilde{\zeta_C}$ [25].

### *2. Calculation of control variables*

More detailed specification on the definition, range, and scaling of the control variable is needed. The details of the definition of control variables in physical space and their dependency on each other are provided here.

*a. Mixture Fraction*

A mixture fraction is defined here following the work of Bilger et al.[55]:

$$Z_1 = \frac{Y_e - Y_e^{Ox}}{Y_e^F - Y_e^{Ox}}, \tag{28}$$

$$Y_e = 2\frac{Y_C}{M_{w,C}} + 0.5\frac{Y_H}{M_{w,H}} - \frac{Y_O}{M_{w,O}}. \tag{29}$$

Here, $Y_e$ and $M_w$ are the element mass fraction and molecular weight that need to be calculated for carbon, hydrogen and oxygen atoms. These values are required to be calculated for fuel and oxidizer for a given boundary condition. This definition of mixture fraction is used to convert the flamelet results in physical space obtained using CHEM1D to flamelet profiles in mixture fraction space needed in the tabulation.

*b. Progress Variable*

A (unscaled) progress variable ($Y_c$) may be defined as a sum of species whose formation indicates the completion of a combustion process. For the present work, species – $CO_2$, $H_2O$, and $H_2$ are chosen to resolve the reaction zone sufficiently. Following the previous works[16,41,45], the control variable $Y_c$ is defined as

$$Y_c = \frac{Y_{CO_2}}{M_{CO_2}} + \frac{Y_{H_2O}}{M_{H_2O}} + \frac{Y_{H_2}}{M_{H_2}}. \tag{30}$$

Here, the reciprocal of molar mass $M$ acts as a weighting factor for the mass fraction $Y$ of species $CO_2$, $H_2O$, and $H_2$. The unscaled progress variable is scaled based on the minimum and maximum values of the progress variable. In this case of four control variables, the extreme value of the progress variable will depend on the mixture fraction, enthalpy loss, and the second mixture fraction describing the mixing of the two oxidizer streams:

$$C = \frac{Y_c - Y_c^u(Z_1, \eta, Z_2)}{Y_c^b(Z_1, \eta, Z_2) - Y_c^u(Z_1, \eta, Z_2)}. \tag{31}$$

Here, 'u' and 'b' indicate $Y_c$ values in unburnt and burnt states.

*c. Enthalpy loss*

The adiabatic enthalpy $h_{ad}$ of a fuel-oxidizer mixture can be expressed as:

$$h_{ad} = Z_1 h_f + (1 - Z_1) h_{Ox}. \tag{32}$$

Here, $h_f$ and $h_{Ox}$ are the adiabatic enthalpy of fuel and oxidizer, respectively. In our case the oxidizer is a mixture of two oxidizer streams, as described by the second mixture fraction $Z_2$. Then the adiabatic enthalpy considering the oxidizer composition is given by:

$$h_{ad} = Z_1 h_f + (1-Z_1)Z_2 h_{Ox_1} + (1-Z_1)(1-Z_2)h_{Ox_2}. \tag{33}$$

Here, $h_{Ox_1}$ and $h_{Ox_2}$ are the adiabatic enthalpy of air and hot-coflow, respectively. The normalized enthalpy loss is calculated as:

$$\eta = \frac{h - h_{ad}}{(1-Z_1)(h_{Ox,\eta=1} - h_{Ox,\eta=0})}. \tag{34}$$

The $h_{Ox,\eta=1}$ and $h_{Ox,\eta=1}$ denote the oxidizer enthalpy at adiabatic condition and condition with the highest loss, set by a chosen minimal temperature, respectively. The minimal temperatures are set at 1030K for hot coflow and 268K for air jet.

*d. Oxidizer mixture fraction*

An oxidizer mixture fraction is introduced to define the mass fraction of the two oxidizers in the mix of oxidizers. In the current case, two oxidizers are air and hot-coflow. It is defined in the present case as:

$$Z_2 = \frac{Y_{O_2} - Y_{O_2,HCF}}{Y_{O_2,Air} - Y_{O_2,HCF}}. \tag{35}$$

Every flamelet has a fixed value of $Z_2$ at its oxidizer boundary. It is a non-reacting scalar. Together with $Z_1$, it fully defines the mass present locally that is originating from each oxidizer stream.

*e. Variances of mixture fraction and progress variable*

The scaled variance in the lookup table varies between zero to one and are related to the variance by:

$$\zeta_{Z_1} = \frac{\widetilde{Z''^2}}{\widetilde{Z}(1-\widetilde{Z})}. \tag{36}$$

Similarly, the scaled variance of progress variable is calculated as:

$$\zeta_{PV} = \frac{\widetilde{Y_c''^2}}{\widetilde{Y_c}(1-\widetilde{Y_c})}. \tag{37}$$

### *3. Calculation of resolved independent scalars and their variances*

The control variables are calculated in the turbulent flow simulation either using the transport equations or through the algebraic equations. The equations of the Favre-averaged control variables – mixture fraction $\widetilde{Z}_1$, oxidizer mixture fraction $\widetilde{Z}_2$, unscaled progress variable $\widetilde{Y}_c$, total absolute enthalpy $\tilde{h}$ and variances of mixture fraction $\widetilde{Z''^2}$ and progress variable $\widetilde{Y_c''^2}$ are provided below,

$$\frac{\partial \bar{\rho}\widetilde{Z}_1}{\partial t} + \frac{\partial}{\partial x_j}(\bar{\rho}\widetilde{u}_j \widetilde{Z}_1) = \frac{\partial}{\partial x_j}\left(\bar{\rho}D_k \frac{\partial \widetilde{Z}_1}{\partial x_j}\right) + S_z, \tag{38}$$

$$\frac{\partial \bar{\rho}\widetilde{Z}_2}{\partial t} + \frac{\partial}{\partial x_j}(\bar{\rho}\widetilde{u}_j \widetilde{Z}_2) = \frac{\partial}{\partial x_j}\left(\bar{\rho}D_k \frac{\partial \widetilde{Z}_2}{\partial x_j}\right), \tag{39}$$

$$\frac{\partial \bar{\rho}\widetilde{Y}_c}{\partial t} + \frac{\partial}{\partial x_j}(\bar{\rho}\widetilde{u}_j \widetilde{Y}_c) = \frac{\partial}{\partial x_j}\left(\bar{\rho}D_k \frac{\partial \widetilde{Y}_c}{\partial x_j}\right) + \overline{\dot{\omega}_{PV}}, \tag{40}$$

$$\frac{\partial \bar{\rho}\tilde{h}}{\partial t} + \frac{\partial}{\partial x_j}(\bar{\rho}\widetilde{u}_j \tilde{h}) = \frac{\partial}{\partial x_j}\left(\bar{\rho}D_k \frac{\partial \tilde{h}}{\partial x_j}\right) + \overline{S_h}, \tag{41}$$

$$\frac{\partial \bar{\rho}\widetilde{Z''^2}}{\partial t} + \frac{\partial}{\partial x_j}(\bar{\rho}\widetilde{u}_j \widetilde{Z''^2}) = \frac{\partial}{\partial x_j}\left(\bar{\rho}D_k \frac{\partial \widetilde{Z''^2}}{\partial x_j}\right) + 2\bar{\rho}D_t\left(\frac{\partial \widetilde{Z}}{\partial x_j}\right)^2 - \bar{\rho}\widetilde{\chi}_Z, \tag{42}$$

Here, $\bar{\rho}$ is the mixture density, $\tilde{u}$ is the mixture velocity, $D_k$ is the total diffusivity equal to $D+D_t$, while $D_t$ is the turbulent diffusivity. The turbulent scalar dissipation rate for variance of mixture fraction $\widetilde{\chi}_Z$ is modelled as:

$$\widetilde{\chi}_Z = C_{Zv}\frac{\varepsilon}{k}\widetilde{Z''^2}, \tag{43}$$

Here, $C_{Zv}$ is assumed as 2. Following previous work[25,52], the variance of progress variable is algebraically calculated assuming the equilibrium of the generation and dissipation of scalar variance at sub-grid scale,

$$\widetilde{Y_c''^2} = C_{Y_c v} \Delta^2 \left[ \left( \frac{\partial \widetilde{Y_c}}{\partial x_i} \right)^2 + \frac{Sc_t}{\mu_t} \left( \overline{Y_c \dot{\omega}_{PV}} - \widetilde{Y_c} \overline{\dot{\omega}_{PV}} \right) \right]. \tag{44}$$

The second term accounts for the progress variable source term. The constant $C_{Y_c v}$ is dynamically calculated similar to the Smagorinsky model with dynamic procedure for sub-grid scale stresses by Lilly[57], extended by Pierce and Moin[58] for calculation of $C_{Zv}$ and further for $C_{Y_c v}$ by Ma (Ch.6 of Ref.25). The turbulent Prandtl number and Schmidt number are assumed constant, equal to 0.7[24,27,52].

**F. Proper Orthogonal Decomposition (POD)**

POD is a mathematical tool that breaks down a complex turbulent flow into modal flow structures based on selected flow-field variables. The modal flow structures can be ordered according to their corresponding eigenvalue. They are the orthonormal eigenvectors from which the entire flow field can be reconstructed. Using a reconstruction based on a low number of modes containing a significant part of the variance ('energy'), a low dimensional representation of the system is obtained, often providing a way to understand critical transport phenomena in the original flow. The detailed methodology is specified in the earlier works[59-67].

The choice of field variable for POD depends on the type of study, flow field, and application. Unlike an incompressible flow, analysis of a compressible or reacting flow should include a thermodynamic quantity such as temperature in addition to the velocity vector field[62]. For autoigniting flames, ignition kernels may be studied using $HO_2$, a species playing an important role in the ignition as used in Ref. 68. First, a variable $\mathbf{X}$ is obtained corresponding to the matrix of values of fields (spanning all grid points) in a set of N snapshots[69] and the associated NxN covariance matrix $\mathbf{A}$:

$$\mathbf{A} = \mathbf{X}^T \mathbf{X}, \tag{45}$$

Next the eigenvalue problem of $\mathbf{A}$ is solved, providing eigenvalues $\lambda_k$ and eigenfunctions $\varphi_k$, $1 \leq k \leq N$:

$$\mathbf{A}j_k = \lambda_k j_k \tag{46}$$

The eigenfunctions are ordered from high to low eigenvalue, expressing the energy content of the modes.

The POD modes are the projections of $\mathbf{X}$ on the eigen vectors:

$$\phi_k = \varphi_k \mathbf{X}, \qquad 1 \leq k \leq N \tag{47}$$

The first P<N modes are sufficient to reconstruct the original spatio-temporal data fields, maintaining the spectral energy contained in the first P modes.

$$\mathbf{X}_{reconst} = \sum_{k=1}^{P} b_k \phi_k \tag{48}$$

The coefficients $b_k$ are time dependent, termed as POD time coefficients. These time coefficients are determined by projecting the time dependent fields on the time-independent POD modes,

$$b_k = \Psi^T \mathbf{x}_k \tag{49}$$

Here $\Psi$ is the matrix of POD modes $[\phi_1, \phi_2, \phi_3, ...]$ and $\mathbf{x}$ the fluctuating field of the covariance matrix $\mathbf{X}$ for a $k^{th}$ snapshot. This method is used here to study the development of the ignition kernel and the flame growth. If $\mathbf{X}$ is the velocity $u'_i$, the trace of the covariance matrix represents the averaged energy of the fluctuating LES velocity field,

$$tr(\mathbf{A}) = \langle (u'_i, u'_i) \rangle = \sum_{k=1}^{N} \lambda_k . \tag{50}$$

Here, $\langle \cdot \rangle$ is the time-averaging operator. Moreover, the analysis has also been performed on velocity-temperature field ($u'_i$, $T'$) by forming a covariance matrix[59,51]:

$$tr(\mathbf{A}) = \sum_{k=1}^{N} \lambda_k = \langle (u'_i, u'_i) \rangle + \gamma^2 \langle (T', T') \rangle . \tag{51}$$

The $\gamma^2 = \langle (u'_i, u'_i) \rangle / \langle (T', T') \rangle$ is a coefficient introduced in Ref. 70 to make the two fluctuating fields consistent with each other.

## III. COMPUTATION DOMAIN

### A. Case Description and Geometry

Fig. 3(a) presents a detailed schematic of the burner. The detailed information related to the configuration and setup is provided in the publication on the experimental work by O'Loughlin & Masri[15]. The design consists of a center carrier jet of cold air with a diameter of 4.6 mm, surrounded by the hot outer coflow of 197 mm, issuing combustion products of hydrogen/air flame. The composition of the fuel, air, and hot coflow is mentioned earlier in Table I.

### B. Boundary Conditions

#### 1. Eulerian boundary condition

The details of the three studied methanol flames – Mt2A, Mt2B and Mt2C are mentioned in case of Table I. Since the evaporation of some fuel in the inner jet is reported in the experiments, the effect of evaporated fuel is also studied here for the Mt2C case. The boundary conditions are shown in the schematics of the computational domain in Fig. 3(b). The cylindrical domain of size is 30D x 66D with two grids of 2.16 million and 3.97 million nodes chosen for the grid independence test. The finer mesh is 1.23 times refined in the axial, radial, and azimuthal directions. The cells in the coarse mesh close to the nozzle are approximately 0.110 mm x 0.115 mm x 0.240 in axial, radial, and azimuthal directions, respectively, to resolve the shear layer and the mixing process adequately. The exact boundary condition for inlet and hot-coflow is provided in Table I for all the simulated cases. The case Mt2C-NG is identical to case Mt2C, except for the fact that the gas phase of the central injector stream does not contain fuel vapor.

TABLE I. Dilute methanol spray in vitiated coflow case variations and boundary conditions.

| No. | Case | $T_{jet}(K)$[a] | $T_{coflow}(K)$[a] | $\dot{m}_f/\dot{m}_c$[b] | $I_f$[c] | $Z_{jet}$[d] | $Z_{coflow}$[d] | $Z_{2,jet}$[e] | $Z_{2,coflow}$[e] | $Y_{c,jet}$[f] | $Y_{c,coflow}$[f] |
|---|---|---|---|---|---|---|---|---|---|---|---|
| 1 | Mt2A | 283 | 1430 | 0.225 | 0.018 | 0.018 | 0 | 1 | 0 | 0 | 5.75 |
| 2 | Mt2B | 288 | 1430 | 0.26 | 0.047 | 0.047 | 0 | 1 | 0 | 0 | 5.75 |
| 3 | Mt2C | 288 | 1430 | 0.295 | 0.080 | 0.080 | 0 | 1 | 0 | 0 | 5.75 |
| 4 | Mt2C-NG | 288 | 1430 | 0.295 | 0.0 | 0.0 | 0 | 1 | 0 | 0 | 5.75 |

[a] $T_{jet}$ and $T_{coflow}$ are the temperatures of central jet and hot coflow.
[b] $\dot{m}_f/\dot{m}_c$ is the fuel loading.
[c] $Y_f$ is the vapor fuel fraction in the central jet (by mass).
[d] $Z_{jet}$ and $Z_{coflow}$ are the mixture fraction at jet and hot coflow, respectively.
[e] $Z_{2,jet}$ and $Z_{2,coflow}$ are the oxidizer mixture fraction at jet and hot coflow, respectively.
[f] $Y_{c,jet}$ and $Y_{c,coflow}$ are the progress variable at jet and hot coflow, respectively.

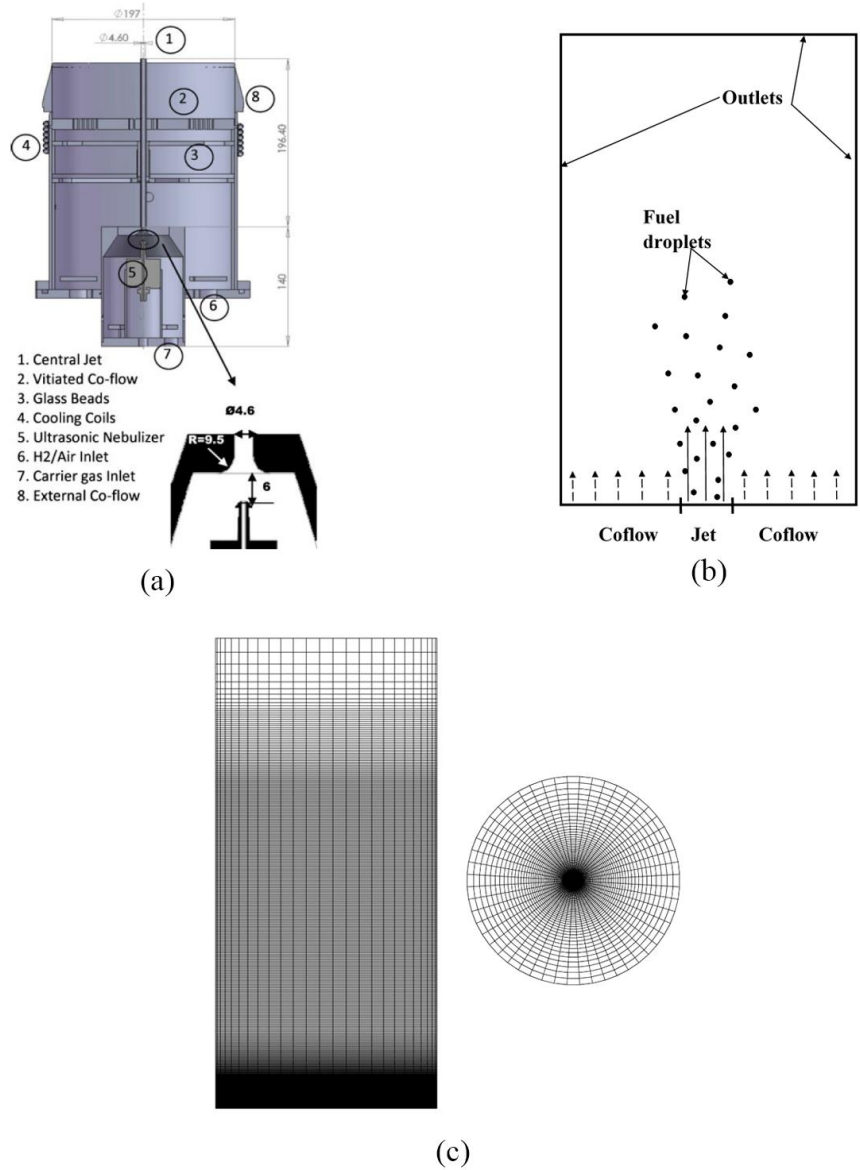

FIG. 3. (a) Schematics of burner for dilute methanol spray in an outer vitiated co-flow[15], (b) schematics of the computational domain, and (c) mesh used in the current study.

### *2. Spray boundary condition*

An essential factor for a dispersed flow is the boundary condition applied to the particles, especially where definite particle statistics and the subsequent dependent processes are required. The auto-igniting dilute spray flames[15] provide extensive data on spray droplet size, mean and fluctuation velocity, temperature, etc., at the inlet. The radial profiles of particle data are provided at the inlet, just above the burner, and at various downstream distances. The spray inlet is divided into multiple injection patches with center points uniformly distributed in the inlet jet-exit plane by dividing its radius into ten intervals and the azimuthal direction into sixteen intervals. The radial divisions are based on the location of experimental measurements. The azimuthal distribution is uniform and sufficiently fine to represent the circumferential uniformity of droplet injection but not too fine to allow the injection of parcels from every patch at each time step.

Further, any patch consists of five overlapping sub-patches to meter the input mass flux of the droplets corresponding to five different size bins. These bins are taken from the experiments – 0 - 10$\mu m$, 10 - 20$\mu m$, 20 – 3re0$\mu m$, 30 - 40$\mu m$ and 40 - 50$\mu m$. In addition to the mass flux, the mean and fluctuating r.m.s. velocity is also fed to injector patches. The data of various radial profiles are provided in a tabulated form for each patch. Hence, multiple droplet parameters are set at the injection. Around 2 million parcels are injected per second from the complete injector at an initial temperature of 283K for Mt2A and 288K for Mt2B and Mt2C cases.

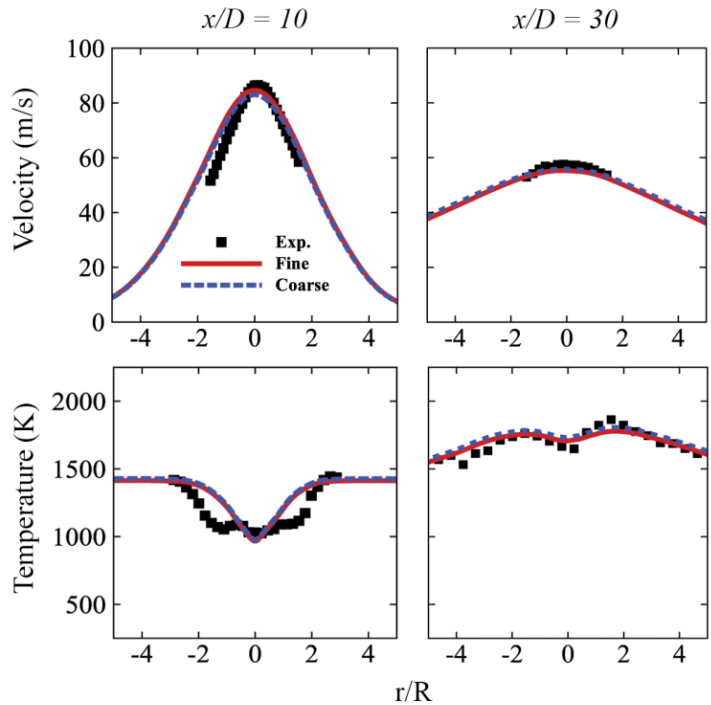

FIG. 4. Comparison of axial velocity and temperature contours at a downstream distance of x/D = 10 and 30 of Mt2C flame for grid independence.

## IV. RESULTS AND DISCUSSION
## A. VALIDATION

### 1. Grid Independence

The coarse mesh out of the two chosen grid sizes has been refined enough to capture essential features of the flow. As mentioned earlier, the mesh resolution close to the jet exit is approximately 0.110 mm x 0.115 mm x 0.240 in axial, radial, and azimuthal directions, respectively. The results of the two grids for the Mt2C flame are compared to check the sensitivity of simulation results to mesh size. The lift-off height, velocity, and temperature distribution are nearly identical on both grids. As representative examples, Fig. 4 reports axial velocity and temperature radial profiles at two downstream locations. The agreement with the experimental data is satisfactory. The grid is sufficiently refined to capture the mixing and reaction zones. Therefore, the coarse mesh is chosen for the rest of the simulations in this study.

Additionally, the grid resolution is also found sufficient according to the LES Index of Resolution Quality (LES_IQ), given by Celik et al.[71]. The recommended value of LES_IQ of more than 75%-80% is acceptable for most cases. The minimum value of LES_IQ is found to be near 95% for both the coarse and fine grid sizes.

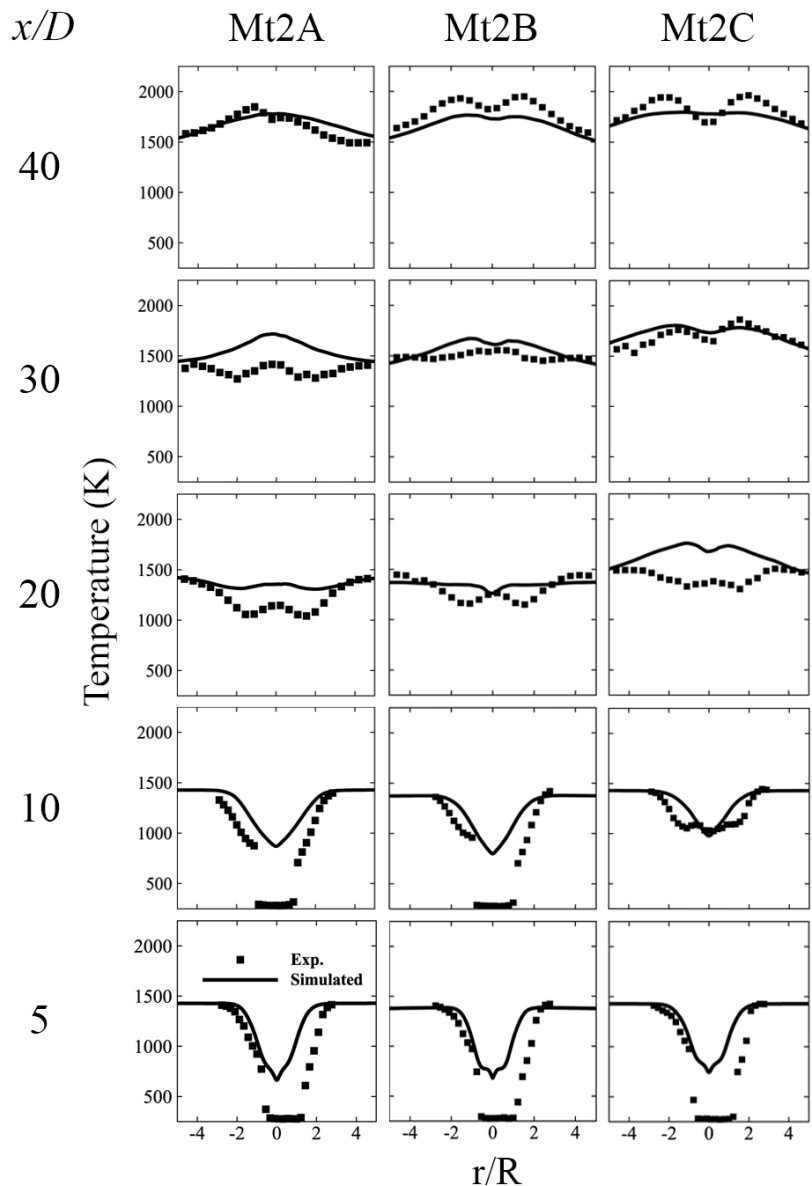

FIG. 5. Comparison of computational and experimental data[15] – radial distribution of mean temperature at different downstream locations of Mt2A, Mt2B, and Mt2C flames.

## *2. Temperature Variation*

The effect of fuel mass loading is noticeable in radial temperature distribution at downstream distances, as shown in Fig. 5. Flame Mt2A, having the lowest fuel loading, has the slowest ignition. This is evident at the downstream distance of 20D from the jet exit, where the temperature of Mt2A is lower than the Mt2C flame in both the experiments and simulation results, albeit there is an over-prediction by the simulations. According to experiments[15], the mean lift-off height of Mt2A lies above 20D, whereas Mt2C undergoes ignition before 10D. This results in a higher temperature at 10D for the Mt2C flame and room temperature for the Mt2A and Mt2B flames, as discernible from experimental data. Although the simulated temperature predictions for Mt2A and Mt2B are lower than Mt2C, large temperature over-predictions are seen for the first two flames. As observed in Ref. 18, there is a large under prediction of temperature for flames Mt2A and Mt2B at x/D=20 or 30, similar to Ref. 20. Similarly, for the Mt2C flame, both studies show large under prediction at x/D=10. In the present study, for the flames Mt2A and Mt2B, the temperatures at the near jet exit locations at x/D=5 and 10 are over predicted, while an over-prediction at x/D=20 occurs for Mt2C. Also, as the temperature of the central jet rises, thickening of the jet-coflow shear layer is observed in the data at x/D=20 for Mt2A and Mt2B flame and at x/D=10 for Mt2C.

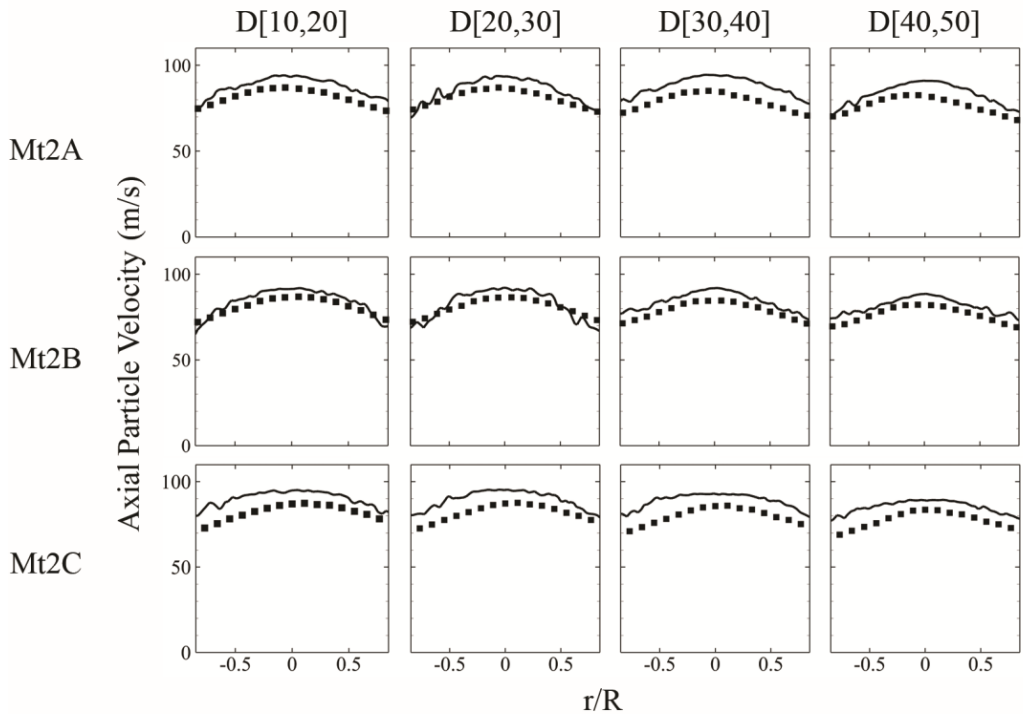

FIG. 6. Comparison of computational (solid lines) and experimental data (symbols)[15] – radial distribution of axial particle velocity at downstream location x/D = 10 of the three flames.

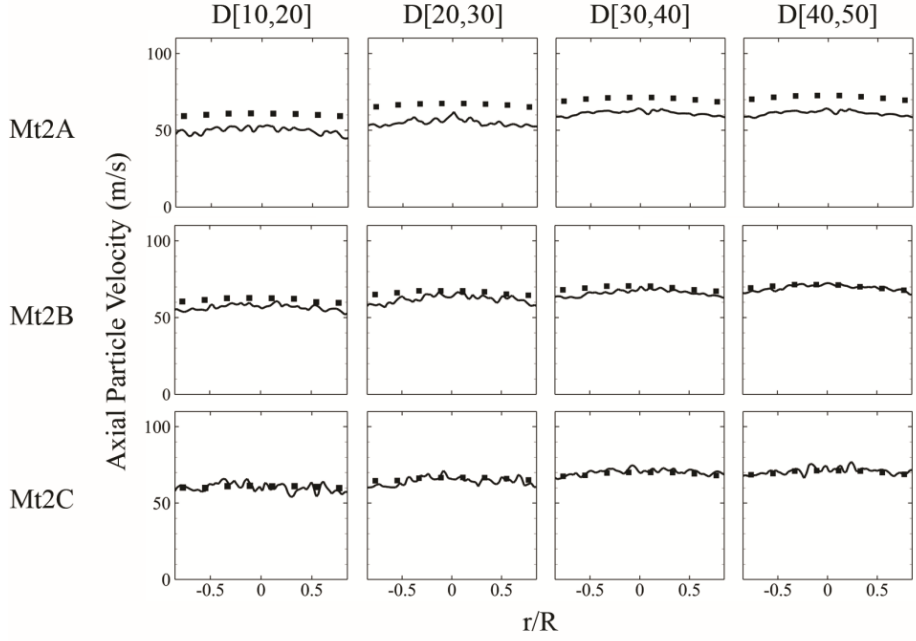

FIG. 7. Comparison of computational (solid lines) and experimental data (symbols)[15] – radial distribution of axial particle velocity at downstream location x/D = 30 of the three flames.

The discrepancy in numerical predictions may be explained in two ways: first, early ignition is predicted at a distance less than 10D in all the three flames, and second, faster mixing of the free shear layer from either side of the jet. As discussed in the following sub-sections, the first is not noticed from the OH distribution, and the ignition locations change with the fuel loading. In contrast, the shear layer development is sensitive to turbulence characteristics. In the present simulations, a single turbulent kinetic energy equation-based model (dynamic *k*-equation) is used, as shown by Eq. 5. The inlet mean velocity and k profile is taken from the experiments[15]. Studying the influence of turbulent inlet modeling or the turbulence model on the flame characteristics near the jet exit may provide insight into this discrepancy.

### *3. Particle Statistics*

Fig. 6 and Fig. 7 show experimental and computed profiles of the mean axial velocity of droplets for different bin sizes for all three flames – Mt2A, Mt2B, and Mt2C at the representative axial distances of 10D and 30D. The radial profile of droplet velocity belonging to a particular size bin is almost constant across all three flames at both downstream locations. The simulation results in several cases over predict the experimental data, namely for size bin D[30,40] in flame Mt2A and all the size bins of Mt2C as shown in Fig. 6. The droplet data of the Mt2A flame at 30D are under-predicted, but the rest is in good agreement for all three flames. It should be noted that there is some asymmetry between the data from the two sides of the central axis, seen in both the experimental and the simulation results. Such asymmetry can result from insufficient statistics in the simulation, but it may have another reason in the experiments. For this reason, a perfect agreement between model and experiment, therefore, cannot be expected. In the flame Mt2A, the biggest droplets have lesser velocity than the small droplets at 10D. But this trend is reversed at a 30D downstream location for all the flames. At 30D, however, the droplet velocity is higher for larger droplets than the smaller ones. This is a consequence of the difference in inertia between small and large droplets. The rapid relaxation of the velocity of small droplets (size less than 10$\mu$m) to gas velocity allows them to be considered as seeding particles for gas velocity measurement[15]. On the hand, the higher inertia of larger droplets delays their response to the velocity changes in the carrier medium. Therefore, at 30D of Mt2C flame, the velocity of droplets with a size less than 20$\mu$m in Fig.8 is closer to the gas velocity in Fig. 4(a) when compared to the droplets of larger size. This also explains a near parabolic profile at 10D compared to the flatter velocity profiles of droplets at 30D downstream from the jet exit.

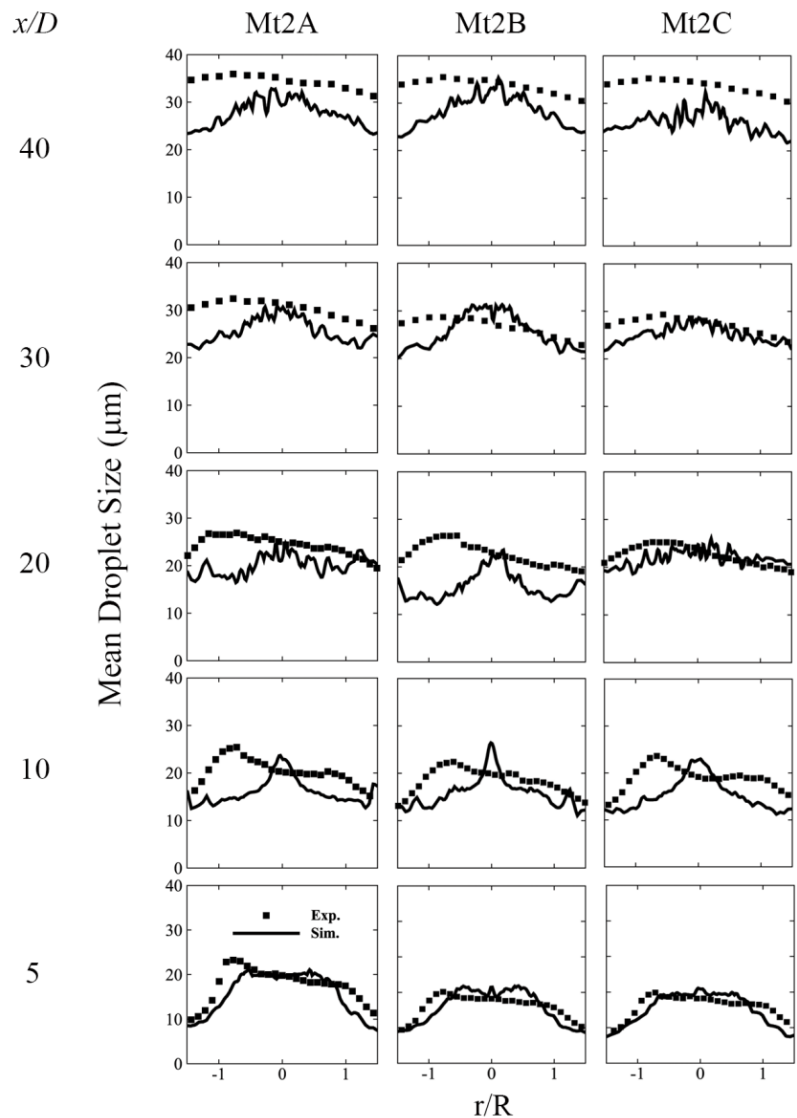

FIG. 8. Comparison of computational and experimental data[15] – radial distribution of mean droplet size at all downstream locations for the three flames.

The droplet mean size is shown in Fig. 8 at various downstream distances for all three flames. It is to be noticed that the droplet size remains approximately constant across all flames at a given downstream location. The droplet size is observed to increase with the downstream location away from the jet exit. The rough portion is observed mainly at 40D, and a smoother curve is obtained at near-jet exit locations. This is because significantly fewer large droplets are available as a sample at 40D. The high temperature of coflow surrounding the center cold-carrier jet evaporates a large number of smaller droplets leaving behind the bigger droplets. Hence, at locations far away from the jet, the region is dominated by fewer but bigger droplets. Except for a few locations, the simulation can adequately capture the details of all three flames.

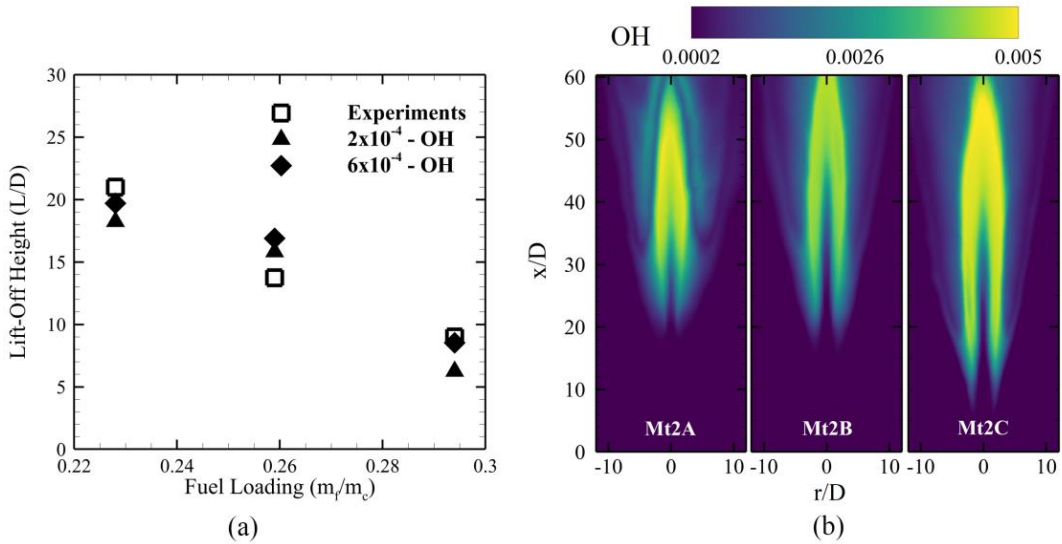

FIG. 9. (a) Lift-off height for minimum values OH mass fraction – $2 \times 10^{-4}$ and $6 \times 10^{-4}$, (b) comparison of OH distribution for flames Mt2A, Mt2B, and Mt2C.

### *4. Lift-off Height and Flame Structure*

As discussed in the previous sub-sections, the flame lift-off height (L) increases on decreasing the fuel loading, as shown in Fig. 9(a). The OH radical is a necessary species for ignition, and its concentration is a reasonable indicator of the auto-ignition as has been used previously[15,28]. Thus the OH concentration is shown in Fig. 9(b). The lift-off height would vary with the chosen minimum threshold value of the OH mass fraction. It was assumed as $2 \times 10^{-4}$ in Refs. 20,72 and $6 \times 10^{-4}$ in Refs. 73-74. Also, the experimental lift-off height is based on visual observation using the still images from a regular digital camera, making it challenging to assume an appropriate OH species mass fraction value for precise validation. Hence, two data sets are considered based on Favre-averaged mean mass fraction of OH as $2 \times 10^{-4}$ and $6 \times 10^{-4}$, as shown in Fig. 9(a). Noticeably, the mean mass fraction of OH value of $6 \times 10^{-4}$ lies near the experimental data for flames Mt2A and Mt2C.

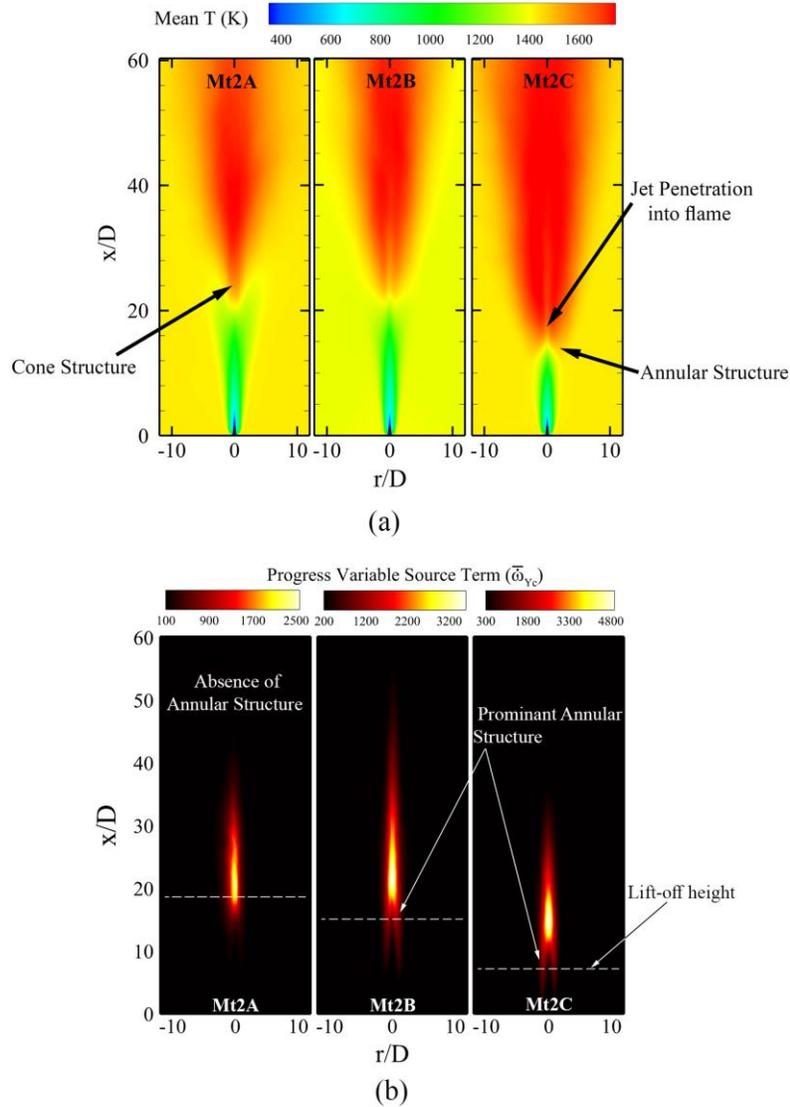

FIG. 10. (a) Temperature distribution for three flames – Mt2A, Mt2B and Mt2C. The flame structure varies with fuel loading – 'cone' type structure of Mt2A and 'annular' structure of Mt2C flame; (b) The mean chemical source term of progress variable equation for three flames. The white-dashed line represents the lift-off height (corresponding to a $2 \times 10^{-4}$ as a mass fraction of OH).

Fig. 10(a) shows the mean temperature contour for the three flames. The basic flame structure is precisely the same as observed in experiments[15] – a 'cone' type structure for Mt2A flame and the 'annular' flame base for Mt2C flame. As reported in experiments, a cone-type structure of flame Mt2A is formed just downstream of the ignition region. The annular structure of Mt2C is just the opposite, with the cold air-fuel jet penetrating the center of the cone of the flame. The chemical source term for the progress variable better represents the presence of the reactions. As shown in Fig. 10(b), the prominent annular structure is visible for Mt2C and Mt2B flame, whereas the annular structure is almost absent in Mt2A flame. Another feature to note is that the white-dashed line representing the lift-off height from OH concentration lies downstream of the chemical source term. It indicates an early start of reactions than predicted by the flame visualization from OH concentration. For flames Mt2B and Mt2C, the annular structure extends well beyond the lift-off height, which is not the case for Mt2A flame. The higher fuel loading results in a near-stoichiometric mixture for flames like Mt2C in the shear layer, initiating the early reactions.

### *5. Effect of evaporated fuel in the carrier jet*

In the additional case of Mt2C-NG, a simulation is carried out without evaporated fuel in the carrier jet. It shows the effect of an already evaporated fuel on the lift-off height of the Mt2C flame, which is around 8 percent of the carrier jet air flow rate or almost one-third of the liquid fuel injected through the nebulizer (refer to Table I). The mean OH distribution is shown for Mt2C (right-side) and Mt2C-NG flame (left-side) in Fig. 11. The OH distribution for Mt2C-NG on the right image indicates an increase of the flame height by a factor of almost five, which is at about 45D downstream of the jet exit. The Mt2C-NG flame is wholly based on the evaporation of liquid spray droplets as the carrier jet mixes with the hot coflow.

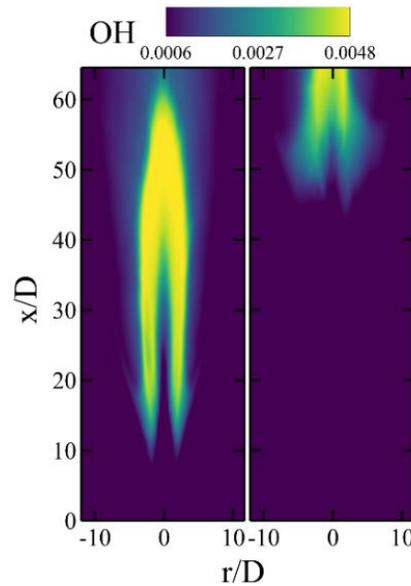

FIG. 11. OH image of Mt2C flame with evaporated fuel (left) and, Mt2C-NG flame without evaporated fuel (right) in carrier jet. Note the high lift-off height in the absence of evaporated fuel.

### B. IGNITION KERNEL FORMATION AND GROWTH

The previous section used OH species as the flame and ignition location indicator. To further study the spread of flame within the shear layer, it is crucial to track the development of OH species in the flow. It is difficult to identify the critical processes in flame evolution and its interaction with flow structures directly from the instantaneous field, especially flame propagation. At any instant, fields like OH species can be split into their mean and fluctuating components. A mean component may provide the lift-off height, as shown in Fig. 9(b), but the fluctuating data helps in the study of flame propagations and ignition kernels. This is addressed using POD.

## 1. Proper Orthogonal Decomposition setup

Proper Orthogonal Decomposition (POD) of a flow identifies the dominant structures, explaining most of the variance of the fluctuations. It allows looking beyond the mean and the insignificant flow features. In Fig. 9(b), the mean distribution of OH primarily lies in the mean shear boundary region. Mixing fuel and oxidizer is vital to sustaining a continuous heat release in the given region. The liquid droplets that evaporate in the neighborhood of hot coflow and the already evaporated fuel from the upstream of a pipe form a premix of fuel and oxidizer in the carrier jet (with very low mixture fraction values). The vortex structures formed in the shear layer between carrier jet and coflow and their transition into turbulence allows the high-temperature oxidizer from vitiated coflow to mix with the fuel in the jet. As discussed in Section II (and also shown in Fig. 2), the flamelets achieve high temperatures at the location near stoichiometric mixture fraction values. The vortices play a significant role in reaching the required mixture fraction value. It may also be pointed out that the value of scalar dissipation rates higher than the critical value prevents the early ignition of fuel with oxidizer from carrier jet or hot vitiated co-flow.

The POD is performed on the fluctuating LES fields of OH and the combined LES fields of velocity and temperature. Around N=200 snapshots of each field – OH, velocity and temperature at a time difference of $2.9011 \times 10^{-5}$ seconds are considered for the analysis. This small time difference is decided based on the signal data from the main simulation and is needed to resolve both the dominant small, turbulent structures and the large vortices. Although the OH forms in the shear layer region on the edges of the carrier jet, as reported in Ref. 15, the scalar POD on the OH field intends to provide the dominant ignition locations and its development in the downstream region. Figure 12(a) shows different POD modes with decreasing order of *species variance* values. *Species variance* refers to the square of the fluctuations of species mass fraction and is analogous to the energy magnitude of velocity fluctuations, also termed fluctuation energy[62,70], which is also evident from Eq. 50. The trace of the covariance matrix of fluctuating OH mass fraction is equal to the time-averaged product of the OH species fluctuations,

$$\text{tr}(\mathbf{A}) = \langle (OH', OH') \rangle = \sum_{k=1}^{N} \lambda_k . \tag{52}$$

Fig. 12(a) shows that the first six modes account for nearly 60 percent of the species variance. The magnitude of this species variance is based on the eigenvalues of the modal decomposition of fluctuating OH field in the spatial domain, whereas the modes are based on the eigenvectors corresponding to the eigenvalues (not shown here for brevity). The number of snapshots considered decides the number of modes (or eigenvalues). The optimum number of "most energetic modes" captures the significant flow features without the rest of the redundant modes with insignificant structures, which helps better understand the flow physics. The POD analysis is carried out on three sets – A, B, and C with 100, 150, and 200 snapshots, respectively, as shown in Fig. 12(a). The first few most energetic modes are resolved for each set, but differences arise in higher modes. The eigenvalue curves of sets B and C nearly overlap, showing that it is sufficient to consider set B for reconstruction and further analysis at a lower computational cost than set C. The fast Fourier transform (FFT) is performed on the POD time coefficients $b_k$, allowing the identification of the frequencies of field fluctuations corresponding to each mode (see Fig. 12(b)) and taking this into account in the analysis of POD results.

### *2. Proper Orthogonal Decomposition results*

The FFT of the first six modes, which constitute 60 percent of the energy (co-variance), shows a distinct peak at around 67.3 Hz and another peak of 269.2 Hz for structures corresponding to the first mode. The second mode also shows a peak at 269.2 Hz, whereas the third mode is again similar to the first mode with the same peaks. The fourth mode shows a peak at 134.4 Hz. It may be noticed that all these frequencies are integer multiples of the base frequency of 67.3 Hz. It indicates the development of periodic OH structures. A better understanding of these dominant structures is achieved when the six most dominant modes are reconstructed for all the snapshots in the physical space-time domain. The location and change of structures of the OH fields may be termed as 'flame propagations' and include the ignition kernels. These features can be investigated in the reconstructed snapshots. Shown in Fig. 12(c) is the six-mode reconstructed 100$^{th}$ snapshot where the (violet-colored) dominant fluctuating OH field can be noticed along with the (yellow-colored) instantaneous vortical structures. The iso-surface value of 8x10$^{-4}$ is chosen for the reconstructed fluctuating OH field, whereas the iso-surface of Q-criterion (at a value equal to 1x10$^8$) is used to show the vortices of the instantaneous field. (The iso-surface of fluctuating OH field at 6x10$^{-4}$ was found to be less suitable for visualization of the OH structures along with vortices).

The instantaneous OH field contains all the structures, unlike the POD reconstructed OH field, which corresponds only to the harmonics of 67.3 Hz. Ignition kernels are the small pockets of reacting mixture formed near the location of lift-off height. The size of kernels may vary from the smallest of the order of filter width to a few mesh cells size. The reconstructed snapshots (as also in Fig. 12(c)) studied do not show the ignition kernel size structures at the flame base. This indicates that the frequency of ignition kernels is not the harmonic of 67.3 Hz. However, a video (added as supplementary material) of OH at a mass fraction of 0.0002 shows a few independent structures forming at the outer periphery of the jet-coflow shear layer, which seems to possess a lower frequency than the smaller, higher frequency ignition kernels at the base of the flame. An instantaneous OH field analysis is more suitable for ignition kernel detection than the POD reconstructed field. Thus, the iso-contour of the instantaneous OH field (yellow) is plotted along with the POD reconstructed field (purple) for the 35$^{th}$, 50$^{th}$, and 100$^{th}$ snapshots in Fig. 13.

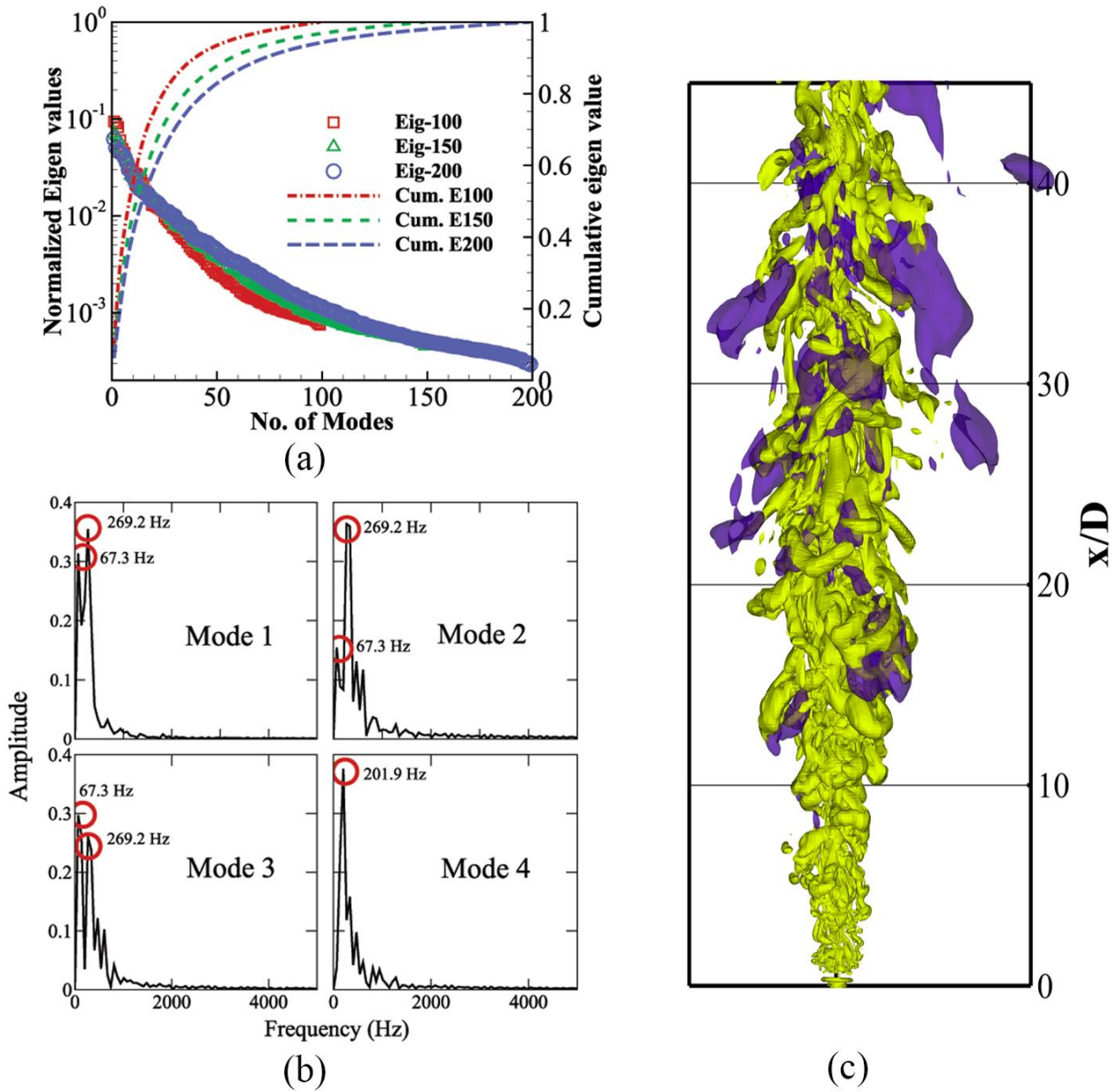

FIG. 12. (a) Normalized eigen values and cumulative energy of the eigenmodes from POD of OH field fluctuations, (b) FFT of the time coefficient corresponding to POD modes, (c) Reconstruction of first six modes of POD showing the development of ignition kernel and flame propagation around instantaneous vortex structures. Q-criterion represents the vortex structures at a value of $1 \times 10^8$ and POD reconstructed OH fluctuation value of $8 \times 10^{-8}$.

The reconstructed field lies within the instantaneous field in most regions for all three snapshots. The iso-contour value of the mass fraction is chosen to be 0.0002, a smaller value for the better visualization of small patches of the field variable associated with ignition kernels. The structures like ignition kernels are verified to be independent and detached from the rest of the OH field by further reducing the value of mass fraction for iso-contour (it is shown here for only 0.0002). The ignition kernels are visible in the instantaneous field, not the reconstructed field. As the flow progresses, these kernels get attached to the rest of the flame or each other through flame propagation (as can be seen in snapshot 100 of Fig. 13). The attachment of ignition kernels happens by forming finger-like streamwise elongated, low-frequency structures. As shown in Fig.13 (and better visualized in the supplementary video file), the ignition kernels are more stable near the shear-layer region and show better flame propagation. On the other hand, the ignition kernels formed near the jet centerline are unstable, and a few of them tend to extinguish as they are convected downstream.

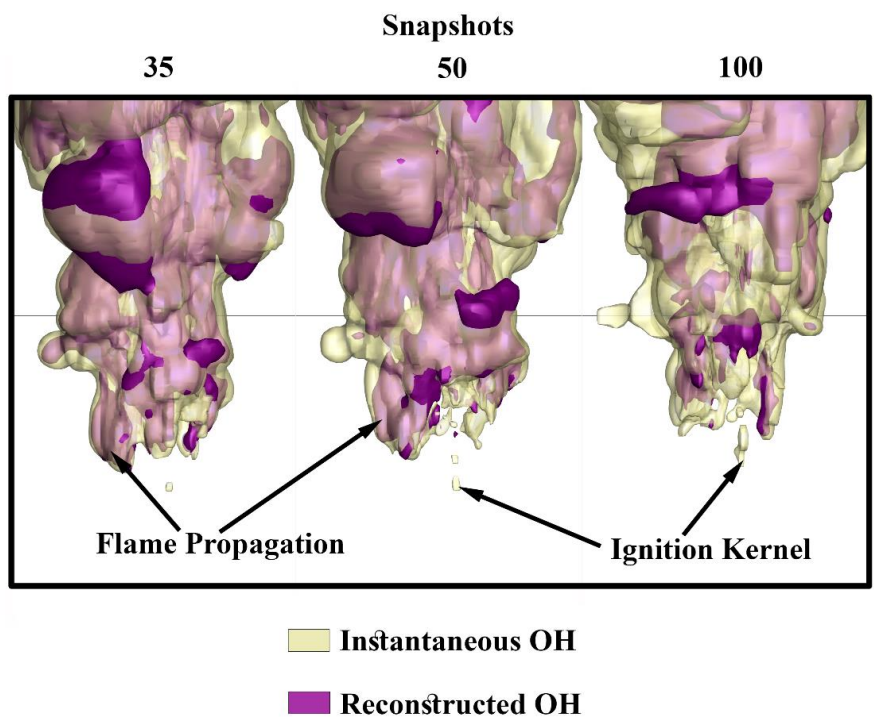

FIG. 13. Iso-contour of the instantaneous field of OH and reconstructed field of six POD modes of fluctuating OH species with the value of $2 \times 10^{-4}$ at three snapshots = 35, 50, and 100. Ignition kernels are visible in the "instantaneous OH" field, whereas the "reconstructed fluctuating OH" field shows the flame propagation. Watch the animation for more details. ( Multimedia video)

On the other hand, flame propagation is analyzed through a reconstructed POD-based fluctuating OH field. A positive iso-contour value of 0.0002 for the fluctuation field represents the flame propagation in the region that favors reactions. Another important observation is that most of the instantaneous field region is occupied by the reconstructed OH field. In other words, the six POD modes are sufficient to resolve the OH field and, thus, capture the flame dynamics. Together, these six POD modes control the flame evolution, whose frequency corresponds to the harmonics of 67.3 Hz (Fig 12(b)). Hence, it can be said that the ignition kernels do not possess frequencies in the harmonics of 67.3 Hz, but the flame propagation, taking place around the kernels and throughout most of the region of flame, does.

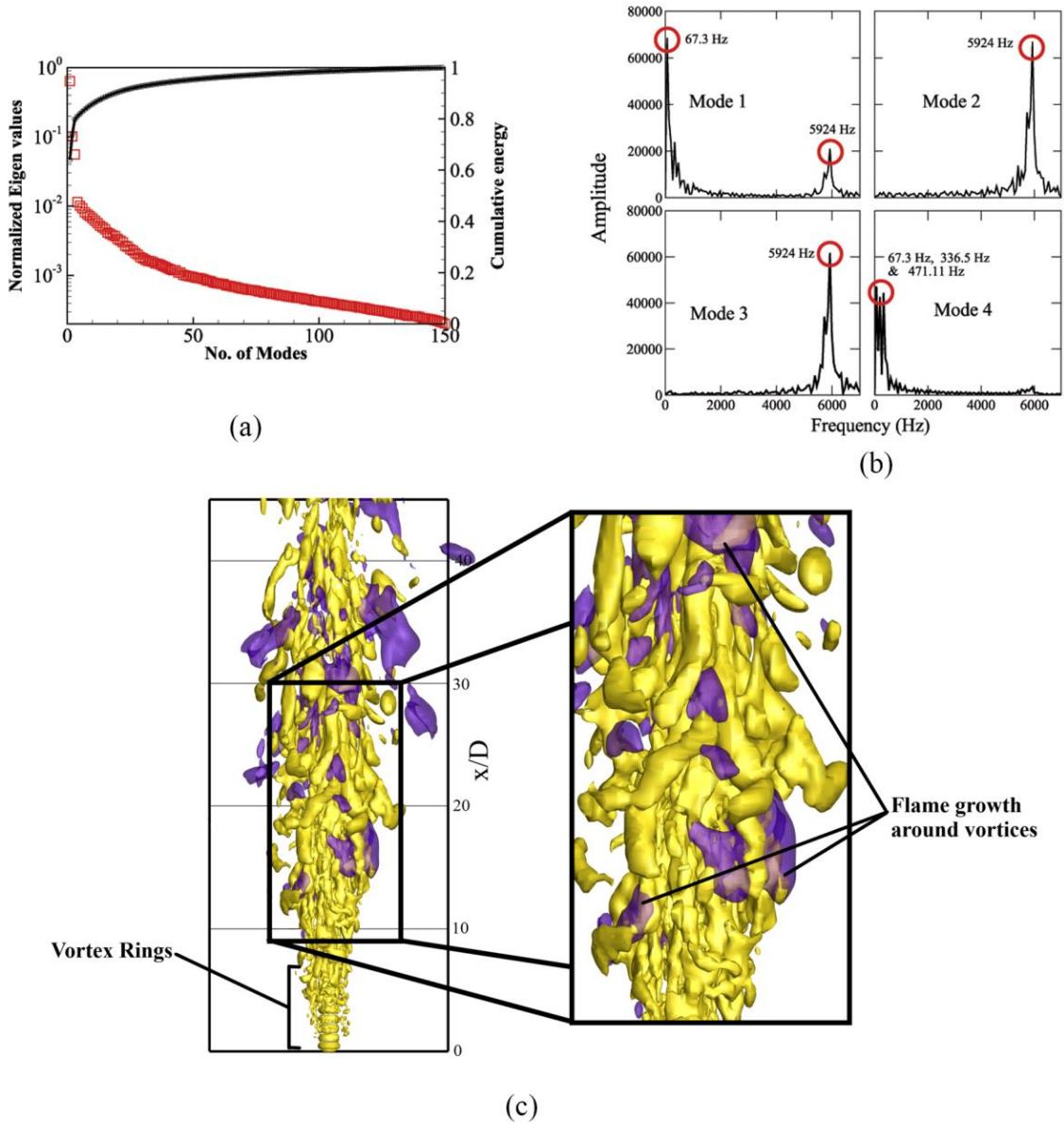

FIG. 14. (a) The energy of the eigenmodes from POD of velocity-temperature field fluctuations, (b) FFT of the time coefficient corresponding to POD modes showing the same frequency as the dominant POD modes of OH fluctuation field, (c) Reconstruction of the first mode of velocity-temperature-based POD showing the development of six-mode reconstructed flame propagation around reconstructed vortex structures. Q-criterion represents the vortex structures at a value of $1 \times 10^8$ and POD reconstructed OH fluctuation value of $8 \times 10^{-8}$. Also, note the vortical ring structures at the bottom along with streamwise vortices, which are part of low-frequency flow structures compared to high-frequency turbulent structures.

In some cases, the flame propagation and development of ignition kernels together play an essential role in the sustenance of the flame at a constant lift-off height. The existence of larger, streamwise tubular structures across the complete circumference at the base of the flame (see Fig. 13) without any ignition kernels upstream to these structures indicates the same (also evident in the supplementary video file). It may only happen if fuel-oxidizer a prior premixing occurs near the jet exit region. Although these streamwise, elongated structures are initially formed due to the flame-propagation of ignition kernels downstream, they manage to sustain at their location because of premix flame-propagation phenomena. The premix flame-propagation is supposed to occur for Mt2A flame for their flat flame base observed in previous research[12]. Here, ignition and flame propagation phenomena compete together in an annular-shaped Mt2C flame. This is only possible when the vaporization and mixing of liquid droplet fuel happen early. In Mt2C flame, the already vaporized fuel at 8 percent is the maximum among the other methanol flames, along with the highest fuel loading, which increases the amount of premixed gas that reaches the appropriate downstream distance before it is auto-ignited.

The conditions that favor the flame propagation have to start with mixing as its first step. This motivates us to perform a velocity-temperature-based POD to assess the vortical structures' contribution to flame propagation. The POD on the velocity-temperature field presents the dominant structures based on the fluctuation energy as denoted in Eq. 51. Since the current case involves high-temperature free-shear flows, we choose the combined velocity-temperature field for POD instead of only velocity-based POD. A previous study by the authors[62] highlights the importance of velocity-temperature-based POD in the evolution of jets in hot-coflow. The first POD mode accounts for approximately 65 percent of the total fluctuation energy, whereas the second and third possess nearly 10 percent and 5 percent, respectively, as depicted in Fig. 14(a). A FFT on the time coefficient of these dominant modes in Fig. 14(b) reveals that the second and third modes correspond to a higher frequency of approximately 5924 Hz, while the first mode shows two dominant peaks – 67.3 Hz and 5924Hz. A high frequency in the range of 6000 Hz essentially corresponds to the turbulent structures which develop majorly in the downstream region. The low frequency of 67.3 Hz indicates the presence of larger vortices and is the same as obtained for fluctuating OH field-based POD. The amplitude peaks near the dominant peak of 67.3Hz are found to be harmonic. Mode 4, with almost 1 percent energy, only shows the dominant peak of 67.3 Hz along with its higher harmonics.

The comparison of the location of the instantaneous OH field from Fig. 13 with the reconstruction of only the first dominant mode of velocity-temperature-based POD and the reconstruction of six OH field-based POD modes shows that the ignition kernels develop in the region (around 5D to 10 D) where the larger streamwise or the broken ring vortical structures are still present (refer to Fig. 14(c)). The OH engulfs more regions around the vortices as the structures evolve further downstream. The larger OH-based structures at downstream locations can also form due to the flame propagation around the "reactive mixture" in the surrounding region of vortices. As discussed in Ref. 15, patches of local extinction and closures in the flame may also give rise to the OH fluctuation field. The zoomed-in image in Fig. 14(c) shows that the two regions – flame development and vortices overlap. The vortex structures allow easy fluid mixing at a lesser strain, as witnessed in the vortical ring structures and the streamwise vortex pairs in the braid region of free-shear jets[62,75]. It enhances the chances of formation of the reaction zones, thus promoting the flame propagation (or maybe ignition kernels) around vortices. Thus, the streamwise braid vortices formed in the region between 5D to 10D (Fig. 14) help create the elongated, tubular streamwise OH-based structures (Fig. 13).

## V. CONCLUSION

LES simulations of methanol spray flames have been carried out using an extended FGM model. The extended FGM model contains an additional parameter (second mixture fraction) as a key feature, allowing the distinction between the two oxidizers (air and hot coflow) in the studied flames. The model requires different laminar flamelet generation for methanol vapor burner in any mixture of the two oxidizers in the presence of heat loss due to evaporation effects. The solver is validated against an extensive auto-igniting dilute methanol flame database in a vitiated coflow[15]. The spray droplets are tracked using Lagrangian particles while the combustion model and the gaseous flow are solved in the Eulerian phase. The pre-evaporated gaseous fuel present in the carrier jet is also considered.

The droplet statistics, including the particle velocity, size, and mean-field temperature, are verified for all three flames with different fuel mass flow rates. The predicted flame structure is in good agreement with the experimental results. The lift-off height is accurately predicted using OH mean-field data, where it is shown that the height decreases with an increase in fuel loading. This flame height is also compared to the flame without the presence of evaporated fuel in the carrier jet for the case with the highest fuel mass flow rate (Mt2C). It is found that not taking into account the pre-evaporated fuel at the jet exit increases the lift-off height by a factor of five.

Further, the Mt2C flame is investigated using proper orthogonal decomposition (POD), performed on the fluctuating OH field and the velocity-temperature fields. The study of the instantaneous OH field is found to be best for the prediction and for revealing ignition kernels. It includes all relevant structures with different frequencies and energy. The POD reconstruction based on the OH field is best suited for flame propagation studies. The ignition kernels formed in the peripheries of the shear layer are observed to be more stable than those near the jet centerline. The tubular, streamwise elongated structure formation precedes the ignition kernel, especially near the shear layer region. These structures seem to self-sustain through flame-propagation without forming any upstream ignition kernels. It is found that these dominant structures corresponding to both the OH and velocity-temperature data fields have an approximately same base frequency of 67.3Hz. The presence of the harmonics of this base frequency also indicates that these structures have periodic motion. The streamwise, braid vortices captured in the first, energetic POD mode seem to assist the formation of tubular-shaped, elongated flame structures. Based on the POD and instantaneous OH-field analysis, it is proposed that the ignition kernels development and flame propagation are dominant features responsible for sustaining the lift-off height of the annular-shaped Mt2C flame.

**DATA AVAILABILITY**

The data that support the findings of this study are available from the corresponding author upon reasonable request.

**ACKNOWLEDGMENTS**

We (B. Bhatia & A. De) acknowledge the National Supercomputing Mission (NSM) for providing computing resources of 'PARAM Sanganak' at IIT Kanpur, which is implemented by C-DAC and supported by the Ministry of Electronics and Information Technology (MeitY) and Department of Science and Technology (DST), Government of India. Also, we would like to thank the computer center (www.iitk.ac.in/cc) at IIT Kanpur for providing the resources to carry out this work.

**REFERENCES**

[1] Cavaliere, A., & De Joannon, M. "Mild combustion." *Progress in Energy and Combustion science* 30, no. 4 (2004): 329-366.
[2] Perpignan, A. A., Rao, A. G., & Roekaerts, D. J. "Flameless combustion and its potential towards gas turbines." *Progress in Energy and Combustion Science* 69 (2018): 28-62.
[3] Wunning JA, Wunning JG. Flameless oxidation to reduce thermal NO mation. Prog Energy Combust Sci 1997;23:81-94.
[4] Weber, R., Gupta, A. K and Mochida, S.: High Temperature Air Combustion (HiTAC): How it all started for Applications in Industrial Furnaces and Future Prospects, Review Paper, Applied Energy, Vol 278, Nov. 15, 2020, 115551. https://doi.org/10.1016/j.apenergy.2020.115551
[5] Ma, L. and Roekaerts, D. "Structure of spray in hot-diluted coflow flames under different coflow conditions: A numerical study". *Combustion and Flame*, 172 (2016): pp.20-37.
[6] Khalil, A. E., & Gupta, A. K. "Impact of pressure on high intensity colorless distributed combustion." *Fuel* 143 (2015): 334-342.
[7] Feser, J. S., Karyeyen, S. and Gupta, A. K.: Flowfield Impact on Distributed Combustion in a Swirl Assisted Burner, *Fuel Journal*, Vol. 263, March 2020. https://doi.org/10.1016/j.fuel.2019.116643
[8] Karyeyen, S., Feser, J. S., Jahoda, E, and Gupta, A. K.: Development of Distributed Combustion Index from a Swirl-Assisted Burner, *Applied Energy*, Vol 268, 15 June 2020, 114967, https://doi.org/10.1016/j.apenergy.2020.114967
[9] Khalil, A. E. and Gupta, A.K.: Flame Fluctuations in Oxy-CO2-Methane Mixtures in Swirl Assisted Distributed Combustion, *Applied Energy*, Vol. 204, pp. 303-317, Oct. 15, 2017. https://doi.org/10.1016/j.apenergy.2017.07.037
[10] Khalil, A. E. and Gupta, A. K.: The Role of $CO_2$ on Oxy-Colorless Distributed Combustion, Applied Energy, Volume 188, February 15, 2017, pp. 466-474. http://dx.doi.org/10.1016/j.apenergy.2016.12.048
[11] Khalil, A. E. and Gupta, A. K.: Towards Colorless Distributed Combustion Regime, *Fuel Journal*, Vol 195, May 2017, pp. 113-122. https://dx.doi.org/10.1016/j.fuel.2016.12.093
[12] Tsuji, H., Gupta, A.K., Hasegawa, T., Katsuki, K., Kishimoto, K. and Morita, M.. "High Temperature Air Combustion-from energy conservation to pollution reduction", by, CRC Press, 2003.
[13] Arghode, V., Gupta, A. K. and Bryden, K.M.: High Intensity Colorless Distributed Combustion for Ultra Low Emissions and Enhanced Performance, J. Applied Energy, Vol. 92, 2012, pp. 822-830. http://doi.org/10.1016/j.apenergy.2011.08.039
[14] Ihme, M., Zhang, J., He, G., and Dally, B. "Large-eddy simulation of a jet-in-hot-coflow burner operating in the oxygen-diluted combustion regime." *Flow, turbulence and combustion* 89, no. 3 (2012): 449-464.
[15] O'Loughlin, W., & Masri, A. R. "The structure of the auto-ignition region of turbulent dilute methanol sprays issuing in a vitiated co-flow." *Flow, turbulence and combustion* 89, no. 1 (2012): 13-35.
[16] Kaul, C. M., Raman, V., Knudsen, E., Richardson, E. S., & Chen, J. H. "Large eddy simulation of a lifted ethylene flame using a dynamic nonequilibrium model for subfilter scalar variance and dissipation rate." *Proceedings of the Combustion Institute* 34, no. 1 (2013): 1289-1297.
[17] Schroll, P., Wandel, A. P., Cant, R. S., & Mastorakos, E. "Direct numerical simulations of autoignition in turbulent two-phase flows." *Proceedings of the Combustion Institute* 32, no. 2 (2009): 2275-2282.

[18]Prasad, V. N., Masri, A. R., Navarro-Martinez, S., & Luo, K. H. "Investigation of auto-ignition in turbulent methanol spray flames using Large Eddy Simulation." *Combustion and flame* 160, no. 12 (2013): 2941-2954.

[19]Heye, C., Raman, V., & Masri, A. R. "Influence of spray/combustion interactions on auto-ignition of methanol spray flames." *Proceedings of the Combustion Institute* 35, no. 2 (2015): 1639-1648.

[20]Sharma, E., De, S., & Cleary, M. J. "LES of a lifted methanol spray flame series using the sparse Lagrangian MMC approach." *Proceedings of the Combustion Institute* 38, no. 2 (2021): 3399-3407.

[21] Lewandowski, M. T., Li, Z., Parente, A., & Pozorski, J. "Generalised Eddy Dissipation Concept for MILD combustion regime at low local Reynolds and Damköhler numbers. Part 2: Validation of the model." *Fuel* 278 (2020): 117773.

[22]De, A., Oldenhof, E., Sathiah, P., & Roekaerts, D. "Numerical simulation of delft-jet-in-hot-coflow (djhc) flames using the eddy dissipation concept model for turbulence–chemistry interaction." *Flow, Turbulence and Combustion* 87, no. 4 (2011): 537-567.

[23]Christo, F. C., & Dally, B. B. "Modeling turbulent reacting jets issuing into a hot and diluted coflow." *Combustion and flame* 142, no. 1-2 (2005): 117-129.

[24]Ma, L., & Roekaerts, D. "Modeling of spray jet flame under MILD condition with non-adiabatic FGM and a new conditional droplet injection model." *Combustion and Flame* 165 (2016): 402-423.

[25]Ma, L. "Computational modeling of turbulent spray combustion." (2016).

[26]Sarras, G., Mahmoudi, Y., Arteaga Mendez, L. D., Van Veen, E. H., Tummers, M. J., & Roekaerts, D. J. E. M. (2014). "Modeling of turbulent natural gas and biogas flames of the Delft Jet-in-Hot-Coflow burner: Effects of coflow temperature, fuel temperature and fuel composition on the flame lift-off height." *Flow, turbulence and combustion* 93, no. 4 (2014): 607-635.

[27]Huang, X., Tummers, M. J., van Veen, E. H., & Roekaerts, D. J. "Modelling of MILD combustion in a lab-scale furnace with an extended FGM model including turbulence–radiation interaction." *Combustion and Flame* 237 (2022): 111634.

[28]Prabasena, B., Röder, M., Kathrotia, T., Riedel, U., Dreier, T., & Schulz, C. "Strain rate and fuel composition dependence of chemiluminescent species profiles in non-premixed counterflow flames: comparison with model results." *Applied Physics B* 107, no. 3 (2012): 561-569.

[29]Blocquet, M., Schoemaecker, C., Amedro, D., Herbinet, O., Battin-Leclerc, F., & Fittschen, C. "Quantification of OH and HO2 radicals during the low-temperature oxidation of hydrocarbons by Fluorescence Assay by Gas Expansion technique." *Proceedings of the National Academy of Sciences* 110, no. 50 (2013): 20014-20017.

[30]Escudero, F., Fuentes, A., Demarco, R., Consalvi, J. L., Liu, F., Elicer-Cortés, J. C., & Fernandez-Pello, C. "Effects of oxygen index on soot production and temperature in an ethylene inverse diffusion flame." *Experimental Thermal and Fluid Science* 73 (2016): 101-108.

[31]Kim, W. W., & Menon, S. "A new dynamic one-equation subgrid-scale model for large eddy simulations." In *33rd Aerospace Sciences Meeting and Exhibit*, p. 356. 1995.

[32]Germano, M., Piomelli, U., Moin, P., & Cabot, W. H. "A dynamic subgrid-scale eddy viscosity model." *Physics of Fluids A: Fluid Dynamics* 3, no. 7 (1991): 1760-1765.

[33]Chai, X., & Mahesh, K. "Dynamic-equation model for large-eddy simulation of compressible flows." *Journal of Fluid Mechanics* 699 (2012): 385-413.

[34]Huang, S., & Li, Q. S. "A new dynamic one-equation subgrid-scale model for large eddy simulations." *International Journal for Numerical Methods in Engineering* 81, no. 7 (2010): 835-865.

[35]Greenshields, C.J. "OpenFOAM: the open source CFD toolbox." *User Guide* (2015).

[36]Schiller, L. "A drag coefficient correlation." *Zeit. Ver. Deutsch. Ing.* 77 (1933): 318-320.

[37]Su, T. F., Patterson, M.A., Reitz, R.D., & Farrell, P. V. "Experimental and numerical studies of high pressure multiple injection sprays." *SAE transactions* (1996): 1281-1292.

[38]Ricart, L. M., Xin, J., Bower, G.R., & Reitz, R.D. "In-cylinder measurement and modeling of liquid fuel spray penetration in a heavy-duty diesel engine." *SAE transactions* (1997): 1622-1640.

[39]Beale, J.C., & Reitz, R.D. "Modeling spray atomization with the Kelvin-Helmholtz/Rayleigh-Taylor hybrid model." *Atomization and sprays* 9, no. 6 (1999).

[40]Ranz, W. R. "Evaporation from drops Part II." *Chem. Eng. Prog.* 48 (1952): 173.

[41]Frössling, N. "The Evaporating of Falling Drops (in German)". *Gerlands Beitrage zur Geophysik*, 52:170–216. (1938).

[42]Abramzon, B., and William A. Sirignano. "Droplet vaporization model for spray combustion calculations." *International journal of heat and mass transfer* 32, no. 9 (1989): 1605-1618.

[43]Zuo, B., Gomes, A. M., & Rutland, C. J. "Modelling superheated fuel sprays and vaproization." *International Journal of Engine*

[44]Daubert, T. E., & Danner, R. P. "Physical and thermodynamic properties of pure chemicals: data compilation, Hemisphere Pub." *Corp., New York* (1989).

[45]Peters, N. "Turbulent combustion." (2001): 2022.

[46]Hermanns, R. T. E. "CHEM1D, a one-dimensional laminar flame code." *Report, Eindhoven University of Technology* (2001).

[47] Ramaekers, W. "Development of flamelet generated manifolds for partially premixed flame simulations", Ph.D. thesis, Technische Universiteit Eindhoven, Eindhoven (2011).

[48] Lindstedt, R. P., & Meyer, M. P. "A dimensionally reduced reaction mechanism for methanol oxidation." *Proceedings of the combustion institute* 29, no. 1 (2002): 1395-1402.

[49] Lamouroux, J., Ihme, M., Fiorina, B., & Gicquel, O. "Tabulated chemistry approach for diluted combustion regimes with internal recirculation and heat losses." *Combustion and flame* 161, no. 8 (2014): 2120-2136.

[50] Van Oijen, J. A., & De Goey, L. P. H. "A numerical study of confined triple flames using a flamelet-generated manifold." *Combustion Theory and Modelling* 8, no. 1 (2004): 141.

[51] Carbonell, D., Oliva, A., & Perez-Segarra, C. D. "Implementation of two-equation soot flamelet models for laminar diffusion flames." *Combustion and Flame* 156, no. 3 (2009): 621-632.

[52] Chrigui, M., Gounder, J., Sadiki, A., Masri, A. R., & Janicka, J. "Partially premixed reacting acetone spray using LES and FGM tabulated chemistry." *Combustion and flame* 159, no. 8 (2012): 2718-2741.

[53] Marracino, B., & Lentini, D. "Radiation modelling in non-luminous nonpremixed turbulent flames." *Combustion science and technology* 128, no. 1-6 (1997): 23-48.

[54] Wang, H., Luo, K., & Fan, J. "Direct numerical simulation and CMC (conditional moment closure) sub-model validation of spray combustion." *Energy* 46, no. 1 (2012): 606-617.

[55] Bilger, R. W. "Turbulent diffusion flames." *Annual Review of Fluid Mechanics* 21, no. 1 (1989): 101-135.

[56] Ma, L., Naud, B., & Roekaerts, D. "Transported PDF modeling of ethanol spray in hot-diluted coflow flame." *Flow, Turbulence and Combustion* 96, no. 2 (2016): 469-502.

[57] Lilly, D. K. "A proposed modification of the Germano subgrid-scale closure method." *Physics of Fluids A: Fluid Dynamics* 4, no. 3 (1992): 633-635.

[58] Pierce, C. D., & Moin, P. "A dynamic model for subgrid-scale variance and dissipation rate of a conserved scalar." *Physics of Fluids* 1

[59] Kosambi, D. D. "Statistics in function space." In *DD Kosambi*, pp. 115-123. Springer, New Delhi, (2016).

[60] Lumley, J. L. "Similarity and the turbulent energy spectrum." *The physics of fluids* 10, no. 4 (1967): 855-858.

[61] Soni, R. K., & De, A. "Investigation of mixing characteristics in strut injectors using modal decomposition." *Physics of Fluids* 30, no. 1 (2018): 016108.

[62] Bhatia, B., & De, A. "Numerical study of trailing and leading vortex dynamics in a forced jet with coflow." *Computers & Fluids* 181 (2019): 314-344.

[63] Soni, R.K., Arya N., and De A. "Modal decomposition of turbulent supersonic cavity." *Shock Waves* 29, no. 1 (2019): 135-151.

[64] Das, P., and De A. "Numerical study of flow physics in supersonic base-flow with mass bleed." *Aerospace Science and Technology* 58 (2016): 1-17.

[65] Das, P., and De A. "Numerical investigation of flow structures around a cylindrical afterbody under supersonic condition." *Aerospace Science and Technology* 47 (2015): 195-209.

[66] Kumar, G., De A., and Harish G.. "Investigation of flow structures in a turbulent separating flow using hybrid RANS-LES model." *International Journal of Numerical Methods for Heat & Fluid Flow* (2017).

[67] Verma, M., Mishra A., and De A. "Flow characteristics of elastically mounted slit cylinder at sub-critical Reynolds number." *Physics of Fluids* 33, no. 12 (2021): 123612.

[68] Danby, S. J., & Echekki, T. "Proper orthogonal decomposition analysis of autoignition simulation data of nonhomogeneous hydrogen–air mixtures." *Combustion and flame* 144, no. 1-2 (2006): 126-138.

[69] Sirovich, L. "Turbulence and the dynamics of coherent structures. I. Coherent structures." *Quarterly of applied mathematics* 45, no. 3 (1987): 561-571.

[70] Puragliesi, R., and Leriche, E. "Proper orthogonal decomposition of a fully confined cubical differentially heated cavity flow at Rayleigh number Ra= $10^9$." *Computers & fluids* 61 (2012): 14-20.

[71] Celik, I. B., Cehreli, Z. N., & Yavuz, I. "Index of resolution quality for large eddy simulations." (2005): 949-958.

[72] Cao, R. R., Pope, S. B., & Masri, A. R. "Turbulent lifted flames in a vitiated coflow investigated using joint PDF calculations." *Combustion and Flame* 142, no. 4 (2005): 438-453.

[73] De, S., De, A., Jaiswal, A., & Dash, A. "Stabilization of lifted hydrogen jet diffusion flame in a vitiated co-flow: Effects of jet and coflow velocities, coflow temperature and mixing." *International journal of hydrogen energy* 41, no. 33 (2016): 15026-15042.

[74] Cabra, R., Myhrvold, T., Chen, J. Y., Dibble, R. W., Karpetis, A. N., & Barlow, R. S. ("Simultaneous laser Raman-Rayleigh-LIF measurements and numerical modeling results of a lifted turbulent H2/N2 jet flame in a vitiated coflow." *Proceedings of the Combustion Institute* 29, no. 2 (2002): 1881-1888.

[75] Hussain, A. F. "Coherent structures and turbulence." *Journal of Fluid Mechanics* 173 (1986): 303-356.